\newif\ifPrintLetter
\newif\ifPrintSupplement
\PrintLettertrue
\PrintSupplementtrue
\ifdefined\LetterOnly\PrintSupplementfalse\fi
\ifdefined\SupplementOnly\PrintLetterfalse\fi
\ifPrintLetter
  \documentclass[aps,prl,reprint,longbibliography,superscriptaddress,nofootinbib,floatfix]{revtex4-2}
\else
  \documentclass[aps,prl,onecolumn,10pt,longbibliography,superscriptaddress,nofootinbib]{revtex4-2}
\fi
\usepackage{amsmath,amssymb,amsthm,bm}
\usepackage{graphicx,booktabs}
\usepackage{xcolor}
\definecolor{linkblue}{RGB}{25,64,102}
\usepackage[colorlinks=true,allcolors=linkblue]{hyperref}
\hypersetup{pdftitle={Magic Can Superactivate Quantum Capacity}}
\makeatletter
\AtBeginDocument{\immediate\write\@auxout{\string\citation{prlTitlesControl}}}
\makeatother
\newcommand{\F}{\mathbb F}
\newcommand{\N}{\mathcal N}
\newcommand{\E}{\mathcal E}
\newcommand{\STAB}{\operatorname{STAB}}
\newcommand{\Tr}{\operatorname{Tr}}
\newcommand{\rank}{\operatorname{rank}}

\newcommand{\ket}[1]{|#1\rangle}
\newcommand{\bra}[1]{\langle#1|}
\newcommand{\proj}[1]{\ket{#1}\!\bra{#1}}
\newcommand{\epsrate}{\varepsilon}
\newcommand{\hb}{h_2}
\newtheoremstyle{prlstatement}{3pt}{3pt}{\normalfont}{}{\itshape}{.---}{.5em}{}
\theoremstyle{prlstatement}
\newtheorem{theorem}{Theorem}

\newtheorem{proposition}{Proposition}

\makeatletter
\renewenvironment{proof}[1][\proofname]{
  \par\pushQED{\qed}\normalfont\topsep6\p@\@plus6\p@\relax
  \trivlist\item[\hskip\labelsep\itshape #1.\textemdash]\ignorespaces
}{\popQED\endtrivlist\@endpefalse}
\makeatother
\usepackage{tikz}
\usetikzlibrary{arrows.meta,positioning,calc}

\newcommand{\PaperAuthors}{%
\author{Chunhe Xiong}
\email{xiongchunhe@csu.edu.cn}
\affiliation{School of Mathematics and Statistics, Central South University, Changsha 410083, China}

\author{Sunho Kim}
\affiliation{School of Mathematical Sciences, Harbin Engineering University, Harbin 150001, China}

\author{Shunlong Luo}
\email{luosl@amt.ac.cn}
\affiliation{Academy of Mathematics and Systems Science, Chinese Academy of Sciences, Beijing 100190, China}
\affiliation{School of Mathematical Sciences, University of Chinese Academy of Sciences, Beijing 100049, China}

\author{Lin Zhang}
\affiliation{School of Mathematical Sciences, Hangzhou Dianzi University, Hangzhou 310018, China}

\author{Shao-Ming Fei}
\email{feishm@cnu.edu.cn}
\affiliation{School of Mathematical Sciences, Capital Normal University, Beijing 100048, China}

\author{Junde Wu}
\email{wjd@zju.edu.cn}
\affiliation{School of Information Engineering, Minnan University of Science and Technology, Quanzhou 362700, China}
\affiliation{School of Mathematical Sciences, Zhejiang University, Hangzhou 310027, China}
}
\begin{document}
\ifPrintLetter
\title{Magic Can Superactivate Quantum Capacity}
\PaperAuthors
\begin{abstract}
Two channels of zero quantum capacity can transmit quantum information
jointly. Environmental magic enables this superactivation in discrete
convolution channels with fixed interactions. We construct a mixed magic
environment for which one channel has positive private but zero quantum
capacity. A second convolution channel, also of zero quantum capacity,
activates it by simulating an erasure channel. Replacing either environment
by any stabilizer mixture makes the joint quantum capacity vanish. The first
environment's modified relative entropy of magic bounds this capacity from
above.
\end{abstract}

\maketitle

Zero quantum capacity need not make a channel useless for quantum
communication. Smith and Yard showed that two channels, each with
zero quantum capacity, can transmit quantum information when used
jointly~\cite{SmithYard2008}. This phenomenon, known as superactivation,
also occurs for Gaussian quantum channels~\cite{SmithSmolinYard2011}
and has no counterpart for the ordinary Shannon capacity of classical
memoryless channels. Related failures of additivity occur for Holevo
information~\cite{Hastings2009} and private capacity~\cite{LiWinterZouGuo2009}.
Recent explicit constructions violate additivity of minimum output
entropy~\cite{ShouGorshkov2026} and exhibit gains in classical
communication from entangled inputs~\cite{ChengLovitzWu2026}.
Quantum capacity can itself be superadditive in simple channel
families~\cite{Leditzky2023,WuZhaoMa2025}. Superactivation goes
further: neither channel alone supports a positive quantum
communication rate, yet together they do.

Discrete quantum convolution offers a way to isolate the role of the
environment in superactivation. A fixed Clifford unitary couples the
input signal to an environment whose initial state determines the
resulting channel~\cite{BGJ2023,BGJ2026}. One resource the environment
can supply is magic, best known for its role in quantum computation
beyond the classically simulable stabilizer
framework~\cite{BravyiKitaev2005,Veitch2014}. Environmental magic can
enhance the capacity of an individual convolution
channel~\cite{BuJaffe2025,SunJin2025}, whereas for the couplings
considered here, any mixture of stabilizer states in the environment
yields zero private and quantum capacities~\cite{Xiong2026}.
Can environmental magic enable two convolution channels to transmit
quantum information together even when each has zero quantum capacity?

To answer this question, we construct a mixed magic environment for
which the resulting convolution channel can transmit classical
information privately at a positive rate but has zero quantum capacity.
By the Smith--Yard construction~\cite{SmithYard2008}, pairing it with
a half-erasure channel gives positive joint quantum capacity. This
auxiliary either transmits its input intact or returns an erasure flag,
with equal probability. We then realize the same mechanism entirely
within the convolution family. Our second convolution channel has zero
quantum capacity but acts as a half-erasure channel on a qubit subspace.
A finite code over multiple uses of this channel simulates an erasure
channel of the required input dimension, with an erasure probability
low enough to activate the first channel. For this pair, magic is
necessary in both environments: with both interactions fixed, replacing
either environment alone by any mixture of stabilizer states reduces
the joint quantum capacity to zero, even if the encoder and decoder
are reoptimized for the new pair. Figure~\ref{fig:setup} summarizes the
two activation routes.

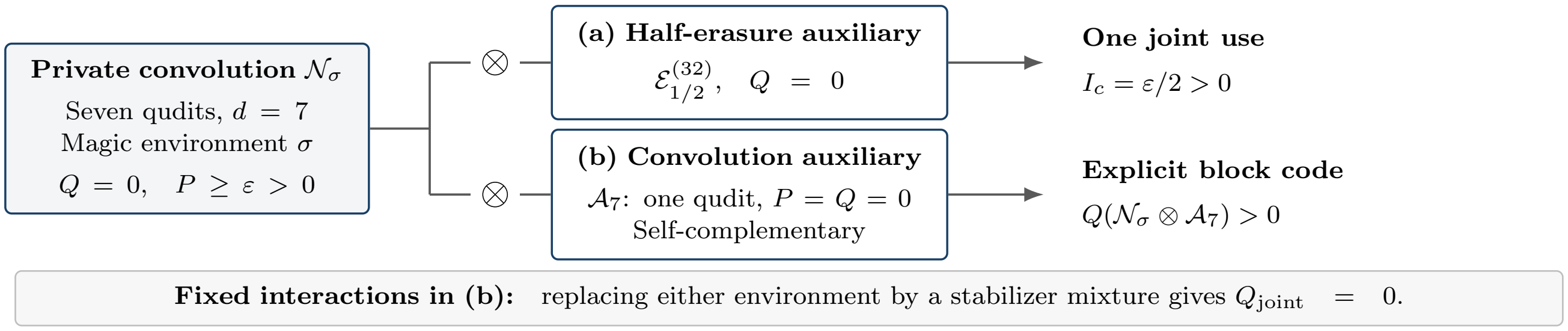
\begin{figure*}[t]
\centering
\begin{tikzpicture}[x=1cm,y=1cm,>=Latex,
 font=\fontsize{8}{10}\selectfont,
 channel/.style={draw=linkblue,rounded corners=2pt,line width=.65pt,
                 align=center,inner sep=5pt},
 helper/.style={channel,text width=4.0cm,minimum height=.95cm},
 result/.style={align=flush left,inner sep=2pt,text width=4.4cm},
 flow/.style={->,line width=.7pt,draw=black!65}]
\path[use as bounding box] (0,0) rectangle (17.55,3.64);

\node[channel,fill=linkblue!5,text width=3.65cm,
      minimum height=1.80cm] (private) at (2.15,2.225)
 {\textbf{Private convolution $\N_\sigma$}\\[3pt]
  Seven qudits, $d=7$\\
  Magic environment $\sigma$\\[3pt]
   $Q=0$,\quad $P\ge\varepsilon>0$};

\node[helper] (erasure) at (8.34,2.95)
 {\textbf{(a) Half-erasure auxiliary}\\[3pt]
  $\mathcal E_{1/2}^{(32)}$,\quad $Q=0$};
\node[helper] (native) at (8.34,1.50)
 {\textbf{(b) Convolution auxiliary}\\[3pt]
  $\mathcal A_7$: one qudit, $P=Q=0$\\
  Self-complementary};

\draw[black!65,line width=.7pt] (private.east) -- (4.82,2.225);
\draw[black!65,line width=.7pt] (4.82,1.50) -- (4.82,2.95);
\draw[black!65,line width=.7pt] (4.82,2.95) -- (erasure.west);
\draw[black!65,line width=.7pt] (4.82,1.50) -- (native.west);
\node[fill=white,inner sep=3pt,font=\fontsize{12}{14}\selectfont]
 at (5.54,2.95) {$\otimes$};
\node[fill=white,inner sep=3pt,font=\fontsize{12}{14}\selectfont]
 at (5.54,1.50) {$\otimes$};

\draw[flow] (erasure.east) -- (11.58,2.95);
\draw[flow] (native.east) -- (11.58,1.50);
\node[result,anchor=west] at (11.92,2.95)
 {\textbf{One joint use}\\[4pt]
  $I_c=\varepsilon/2>0$};
\node[result,anchor=west] at (11.92,1.50)
 {\textbf{Explicit block code}\\[4pt]
  $Q(\N_\sigma\otimes\mathcal A_7)>0$};

\node[draw=black!25,fill=black!3,rounded corners=2pt,line width=.5pt,
      align=center,text width=16.75cm,minimum height=.60cm,
      inner sep=4pt] at (8.775,.36)
 {\textbf{Fixed interactions in (b):}\quad
  replacing either environment by a stabilizer mixture gives $Q_{\rm joint}=0$.};
\end{tikzpicture}
\caption{Two activation routes for the same private convolution channel.
Here $P$ and $Q$ are the private and quantum capacities, and $\varepsilon$
is the private information of the chosen input ensemble.
(a) A 32-dimensional half-erasure channel gives coherent information
$I_c=\varepsilon/2$ in one joint use. (b) A finite code allows a second
convolution channel to activate the first. With both interactions fixed,
replacing either environment by a stabilizer mixture makes the joint
quantum capacity zero, even after reoptimizing the code.
The authors used ChatGPT to help write the TikZ code for this schematic.}
\label{fig:setup}
\end{figure*}

\emph{Setting.---}
Let $d$ be an odd prime, $V=\mathbb F_d^n$, and
$\mathcal H=(\mathbb C^d)^{\otimes n}$, of dimension $D=d^n$.
For $j\in\mathbb F_d$, the single-qudit Pauli generators act as
$X\ket j=\ket{j+1}$ and $Z\ket j=\exp(2\pi i j/d)\ket j$.
Tensor products of their powers, with a choice of phase, are the
Weyl operators. A pure stabilizer state is a common eigenstate of
a maximal commuting family of Weyl operators. The labels of such a
family form a Lagrangian subspace of the symplectic space $V\oplus V$.
We call states in the convex hull $\STAB_n$ of these pure states
stabilizer mixtures, and states outside it magic~\cite{Veitch2014}.
A Clifford unitary preserves the Pauli group under conjugation. The Weyl convention and stabilizer projectors are given in Sec.~SI of the Supplemental Material~\cite{supplement}.

In the discrete convolution of Bu, Gu, and Jaffe
(BGJ)~\cite{BGJ2023,BGJ2026}, a matrix
$G\in\mathrm{GL}(2,\mathbb F_d)$ specifies the interaction.
We consider positive couplings, meaning that all four entries are
nonzero. Writing $G^{-\mathsf T}=(G^{-1})^{\mathsf T}
=\left(\begin{smallmatrix}a&b\\c&e\end{smallmatrix}\right)$,
the Clifford unitary acts on the signal $A$ and environment $Y$ as
\begin{equation}
 U_G\ket{x,y}_{AY}=\ket{ax+by,cx+ey}_{BE},\qquad x,y\in V,
 \label{eq:convolution-unitary}
\end{equation}
where $\ket{x,y}_{AY}=\ket{x}_A\otimes\ket{y}_Y$ and the arithmetic is componentwise over $\mathbb F_d$.
For a fixed environment state $\sigma_Y$ independent of the signal,
Bob receives $\N_{G,\sigma}(\rho)=\rho\boxtimes_G\sigma
:=\Tr_E[U_G(\rho_A\otimes\sigma_Y)U_G^\dagger]$.
Purifying $\sigma_Y$ to $\ket{\Sigma}_{YF}$ gives the isometry
$V_\sigma\ket\psi=(U_G\otimes I_F)(\ket\psi_A\otimes\ket\Sigma_{YF})$.
Eve receives $K=EF$, so the channel complementary to
$\N=\N_{G,\sigma}$ is $\N^c(\rho)=\Tr_B[V_\sigma\rho V_\sigma^\dagger]$.

The unassisted private and quantum capacities $P$ and $Q$ are
the optimal asymptotic rates for private classical and quantum
communication, in bits and qubits per use, respectively. If Alice
chooses $\rho_t$ with probability $p_t$ and records $t$ in a
classical register $T$, her record and the outputs have joint state
\begin{equation}
 \Omega_{TBK}=\sum_t p_t\proj{t}_T\otimes V_\sigma\rho_tV_\sigma^\dagger .
 \label{eq:common-state}
\end{equation}
The private information $\Delta=I(T;B)-I(T;K)$ compares the two
outputs for the same label. Since $I(T;O)=H(T)-H(T|O)$, the
label entropy cancels. The coding bounds~\cite{Devetak2005} are
\begin{equation}
 \begin{aligned}
 P(\N)&\ge\Delta=H(T|K)_\Omega-H(T|B)_\Omega,\\
 Q(\N)&\ge I_c(\rho,\N)=S(\N(\rho))-S(\N^c(\rho)).
 \end{aligned}
 \label{eq:coding-bounds}
\end{equation}
Here $H(T|O)_\Omega=S(\Omega_{TO})-S(\Omega_O)$, and $S$ is the
von Neumann entropy in bits. For a joint input across $\ell$ uses,
the coherent information divided by $\ell$ is again a lower
bound on $Q(\N)$.

\emph{An explicit private convolution channel.---}
Take $d=n=7$ and choose
$G_{\rm priv}=3^{-1}
 \left(\begin{smallmatrix}1&2\\-1&1\end{smallmatrix}\right)$
over $\F_7$. The corresponding interaction $U=U_{G_{\rm priv}}$ is
\begin{equation}
 U\ket{x,y}_{AY}=\ket{x+y,y-2x}_{BE},
~ x,y\in\F_7^7.
 \label{eq:example-interaction}
\end{equation}

We choose the environment so that the Choi state has positive
partial transpose (PPT) and Bob's output carries more information
than Eve's about a binary signal. Define sixteen points in $\F_7^3$ by
\begin{equation}
 a_{4u+v}=(u,v,u^2-3v^2),\qquad u,v\in\{0,1,2,3\}.
 \label{eq:example-points}
\end{equation}
The subscript $4u+v$ is an integer index. Using blocks of sizes
$1,3,3$ and $0_3=(0,0,0)$, define for $0\le i,j\le15$
\begin{equation}
 \begin{aligned}
 r_i&=(1,a_i,0_3),  & s_j&=(0,0_3,a_j),\\
 u_{ij}&=-r_i-3s_j, & v_{ij}&=-3r_i-s_j,\\
 w_i&=-4r_i,        & z_j&=-4s_j.
 \end{aligned}
 \label{eq:example-labels}
\end{equation}
The first row labels the two signals; the other rows label the environment.

Let $H=(H_{ij})=
\left(\begin{smallmatrix}1&1\\1&-1\end{smallmatrix}\right)^{\otimes4}$
be the real Hadamard matrix, so $H_{ij}\in\{+1,-1\}$ and
$HH^{\mathsf T}=16I_{16}$. With the unnormalized vectors
$\ket{\phi_{ij}}=\ket{u_{ij}}+H_{ij}\ket{v_{ij}}$, choose the
environment state $\sigma$ on $Y$ as
\begin{equation}
 \begin{aligned}
 640\,\sigma={}\sum_{i,j}\ket{\phi_{ij}}\bra{\phi_{ij}}+4\sum_i\proj{w_i}+4\sum_j\proj{z_j}.
 \end{aligned}
 \label{eq:environment}
\end{equation}
The four families have distinct first coordinates $6,4,3,0$.
Within each family, the remaining coordinates determine the indices.
Thus $\sigma$ involves
$2\cdot16^2+2\cdot16=544$ distinct computational-basis labels.
Each $\ket{\phi_{ij}}$ has squared norm $2$, giving
$\Tr\sigma=(2\cdot16^2+4\cdot32)/640=1$.
Write $\N_\sigma=\N_{G_{\rm priv},\sigma}$ for the resulting channel.

We first show that $Q(\N_\sigma)=0$. Let $R_A\simeq A$ be a
reference register and let
$\ket{\Phi_D}=D^{-1/2}\sum_{x\in\F_7^7}\ket{x}_{R_A}\ket{x}_A$,
where $D=7^7$. The normalized Choi state is
$J=(\operatorname{id}_{R_A}\otimes\N_\sigma)(\proj{\Phi_D})$.
After partial transpose on $B$, the matrix element
$\langle x,\beta|J^{T_B}|x',\beta'\rangle$ vanishes unless
$\beta+3x=\beta'+3x'$. Thus $s=\beta+3x$ labels the blocks.
Within each block, use $u=\beta-x$ as the basis label. In this
basis the $D$ blocks are identical, each equal to
$\mathcal R_3(\sigma)/D$, where
\[
 [\mathcal R_3(X)]_{u,v}
 =X_{(u+3v)/4,(3u+v)/4},~ u,v\in\F_7^7 .
\]
The block decomposition is derived in the Supplemental
Material~\cite{supplement}.
This entry rearrangement fixes the diagonal. Since
$(w_i+3z_j)/4=u_{ij}$ and $(3w_i+z_j)/4=v_{ij}$, it moves the
$(u_{ij},v_{ij})$ entry to $(w_i,z_j)$ and likewise for its
adjoint. Hence
\begin{equation}
 640\mathcal R_3(\sigma)\simeq I_{512}\oplus
 \begin{pmatrix}4I_{16}&H\\H^{\mathsf T}&4I_{16}\end{pmatrix}
 \oplus 0 ,
 \label{eq:example-ppt}
\end{equation}
where $\simeq$ denotes a simultaneous permutation of rows and
columns. Since $HH^{\mathsf T}=16I_{16}$, all singular values of
$H$ are $4$, and the middle block has eigenvalues $4\pm4$.
Thus $J^{T_B}\succeq0$. A channel with a PPT Choi state has zero quantum
capacity~\cite{Horodecki1998,Horodecki2000Binding}.

To show that $P(\N_\sigma)>0$, let $T=0,1$ label the two
equiprobable signal states
\begin{equation}
 \rho_0=\frac1{16}\sum_i\proj{r_i},~
 \rho_1=\frac1{16}\sum_j\proj{s_j}.
 \label{eq:private-inputs}
\end{equation}
Purify $\sigma$ by giving the two endpoints $u_{ij},v_{ij}$ of
each coherent pair the same flag $ij$ in $F$, and giving each
$w_i$ and $z_j$ its own flag. All distinct flags are orthogonal.
For a signal basis label $x$ and an environment basis label $y$,
Bob's output label is $x+y$ and Eve's is $(y-2x,f(y))$, where
$f(y)$ is the flag of $y$.
For fixed $x$, the maps $y\mapsto x+y$ and $y\mapsto y-2x$ are
both injective, so tracing out $B$ or $EF$ removes all cross terms
between distinct $y$. For each receiver, the two signal states
therefore give outputs diagonal in a common basis: the computational
basis of $B$ for Bob, and the product of the computational basis
of $E$ and the flag basis of $F$ for Eve.

Thus $H(T|B)$ and $H(T|K)$ are averages of the binary posterior
entropy of $T$. Only ambiguous outputs, those possible for both
signals, contribute: all others determine $T$. For each listed
output $o$ of $O=B$ or $K$, the columns give $\Pr(T=t,O=o)$
in units of $1/(2\cdot16\cdot640)$.
\[
\begin{array}{lcc}
 \text{output label} & T=0 & T=1\\ \hline
 B:\ -3s_j                &16&4\\
 B:\ -3r_i                &4&16\\
 K:\ (-3r_i-3s_j,\,ij)   &1&1
\end{array}
\]

For fixed $j$, the signal $\rho_0$ gives Bob the output $-3s_j$
through all sixteen pairs $(r_i,u_{ij})$, each of weight $1$,
whereas $\rho_1$ gives the same output through $s_j+z_j=-3s_j$,
with weight $4$. For the output $-3r_i$, the weights are reversed:
$r_i+w_i=s_j+v_{ij}=-3r_i$ for every $j$. For Eve,
$u_{ij}-2r_i=v_{ij}-2s_j$ gives one contribution of weight $1$
from each signal, both with flag $ij$. The table exhausts the
ambiguous outputs. For Bob, comparing the first coordinates leaves
only the two types of overlap shown in the table. The identity
$a_i+3a_j=4a_\ell\Rightarrow i=j=\ell$ excludes any further
matches within those types. The full enumeration for Bob and
Eve is given in the Supplemental Material~\cite{supplement}.

Bob's $32$ ambiguous outputs have total probability $1/32$ and
posterior odds $4:1$ or $1:4$. Eve's $256$ ambiguous outputs have
total probability $1/40$ and posterior probabilities $1/2,1/2$.
Thus, writing $h_2$ for the binary entropy, we obtain
\begin{equation}
 H(T|B)=\frac{\hb(1/5)}{32},\qquad H(T|K)=\frac1{40}.
 \label{eq:conditional-entropies}
\end{equation}
The private information of this ensemble is therefore
$\varepsilon=1/40-\hb(1/5)/32$. Equation~\eqref{eq:coding-bounds}
and the PPT result give
\begin{equation}
 P(\N_\sigma)\ge\varepsilon
 =\frac{12/5-\log_2 5}{32}>0,\qquad Q(\N_\sigma)=0,
 \label{eq:separation}
\end{equation}
where $5^5<2^{12}$ proves the strict inequality.
For this coupling, every stabilizer mixture induces an
entanglement-breaking channel~\cite{Xiong2026}, whose private
capacity is zero. Hence $P(\N_\sigma)>0$ certifies that the
environment state $\sigma$ is magic.

\emph{Half-erasure activation.---}
Bound entanglement can yield secret key~\cite{Horodecki2005,Horodecki2008,Ozols2014}.
Smith and Yard convert private information into joint coherent
information using half erasure~\cite{SmithYard2008}.
The channel $\E_p^{(m)}$ transmits its $m$-dimensional input with
probability $1-p$ and returns an orthogonal flag otherwise. Its
complement transmits with probability $p$. For $p\ge1/2$, further
erasure reproduces Bob's output, so $\E_p^{(m)}$ is antidegradable
and has zero quantum capacity~\cite{Bennett1997}.

A rank-$r$ state can be purified in $r$ dimensions. For our
two rank-$16$ signals we choose orthogonal sectors $C_0,C_1$, so
$\dim C=16+16=32$. Explicitly,
\begin{equation}
 \begin{aligned}
 \ket{\psi_0}_{AC}&=\tfrac14\sum_i\ket{r_i}_A\ket{0,i}_C,\\
 \ket{\psi_1}_{AC}&=\tfrac14\sum_j\ket{s_j}_A\ket{1,j}_C.
 \end{aligned}
 \label{eq:activation-input}
\end{equation}
Alice prepares $\omega_{AC}=\tfrac12\proj{\psi_0}+\tfrac12\proj{\psi_1}$
and sends $A$ through $\N_\sigma$ and $C$ through $\E_p^{(32)}$.
The transmitted and erased branches give~\cite{supplement}
\begin{equation}
 \begin{aligned}
 f(p)&:=I_c(\omega_{AC},\N_\sigma\otimes\E_p^{(32)})\\
     &=(1-p)H(T|K)-pH(T|B)\\
     &=\frac{1-p}{40}-\frac{p}{32}\hb(1/5).
 \end{aligned}
 \label{eq:erasure-rate}
\end{equation}
At $p=1/2$, $f(1/2)=\varepsilon/2$. More generally, the same
input $\omega_{AC}$ gives $f(p)>0$ while both individual quantum
capacities vanish throughout the interval
\begin{equation}
 \frac12\le p<p_*,
 \qquad p_*:=\frac4{4+5\hb(1/5)}\approx0.525649 .
 \label{eq:erasure-window}
\end{equation}

\emph{General private environments.---}
We now vary both the local dimension and the coupling while keeping
the environment on seven qudits. For a positive coupling, write
$r=ae/(bc)$ for its cross ratio; invertibility gives $r\ne1$.

\begin{theorem}[Magic environments for private PPT channels]
\label{thm:private}
For every prime $d\ge7$ and every positive coupling with
$r\notin\{-1,1/2,1/3\}$, there is a mixed magic environment $\sigma$ on seven qudits
such that $\N_{G,\sigma}$ has a PPT Choi state and
\[
 Q(\N_{G,\sigma})=0,~ P(\N_{G,\sigma})\ge\varepsilon>0.
\]
It also satisfies
$Q(\N_{G,\sigma}\otimes\E_{1/2}^{(32)})\ge\varepsilon/2$.
\end{theorem}

\begin{proof}[Proof sketch]
Rescaling the signal input and the two outputs reduces $U_G$ to
$U_h\ket{x,y}=\ket{x+y,y-hx}$ with $h=-1/r$, leaving the
environment input unchanged. Set $\alpha=1+h$; the hypotheses
give $\alpha\notin\{0,\pm1,\pm2\}$.
Choose a nonsquare $\nu\in\F_d$ and use the sixteen points
$a_{4u+v}=(u,v,u^2-\nu v^2)$, $u,v\in\{0,1,2,3\}$.
In Eq.~\eqref{eq:example-labels}, replace the label coefficients
$3,4$ by $\alpha,\alpha+1$, respectively. The Hadamard matrix
and the real weights in Eq.~\eqref{eq:environment} are unchanged.

Under the Choi rearrangement, the $(u_{ij},v_{ij})$ entry still
moves to $(w_i,z_j)$, while diagonal entries stay fixed. This gives
the same positive semidefinite Hadamard block as in
Eq.~\eqref{eq:example-ppt}, with only nonnegative diagonal entries
outside it. Hence the Choi state is PPT and $Q=0$.

For privacy, the key property of the new point set is
\[
 a_i+\alpha a_j=(1+\alpha)a_\ell
 \quad\Longrightarrow\quad i=j=\ell .
\]
Indeed, writing $a_i=(\xi_i,q_\nu(\xi_i))$ with
$q_\nu(u,v)=u^2-\nu v^2$, the equality forces
$q_\nu(\xi_i-\xi_j)=0$. Since $\nu$ is a nonsquare, this implies
$\xi_i=\xi_j=\xi_\ell$. Together with the exclusions on $\alpha$,
this rules out extra overlaps for Bob. As in the example, Bob has
32 shared outputs with weights $(16,4)$ or $(4,16)$. Eve has one
shared output per pair flag, giving 256 with weights $(1,1)$.
With the same normalization, both entropies in
Eq.~\eqref{eq:conditional-entropies} are unchanged; see Sec.~SIII
of the Supplemental Material~\cite{supplement} for the full calculation.

Undoing the signal basis change preserves the private information
$\varepsilon$. Rescaling $A$ alone gives the joint input
$\omega^G_{AC}$ for $U_G$, with coherent information $f(p)$ for
every $p$. Positive private capacity rules out a stabilizer mixture
for these couplings, so $\sigma$ is magic.
\end{proof}

\emph{A convolution auxiliary.---}
To realize erasure on the five-qubit register $C$, we first identify
a qubit half-erasure sector in a second convolution channel.
For $x,y\in\F_d$, take
\begin{equation}
 U_{\rm aux}\ket{x,y}=\ket{x+y,x-y},~
 \ket{\xi}=\frac{\ket{1}+\ket{-1}}{\sqrt{2}}.
 \label{eq:helper-unitary}
\end{equation}
The BGJ matrix is
$G_{\rm aux}=2^{-1}
 \left(\begin{smallmatrix}1&1\\1&-1\end{smallmatrix}\right)$.
Let $B_h,E_h$ be the two outputs and let $\mathcal A_d$ be the channel
to $B_h$, with isometry
$V_{\rm aux}\ket{\psi}=U_{\rm aux}(\ket{\psi}\otimes\ket{\xi})$.
Swapping the outputs replaces $y$ by $-y$ in
Eq.~\eqref{eq:helper-unitary}. Since $\ket{\xi}$ is invariant under
this reflection, $\mathcal A_d$ equals its complement after
identifying the two output spaces. Thus
$P(\mathcal A_d)=Q(\mathcal A_d)=0$ on the full input space.

On the subspace spanned by $\ket{1}$ and $\ket{-1}$,
\begin{equation}
 V_{\rm aux}\ket{\pm1}
 =\frac{\ket{\pm2}_{B_h}\ket{0}_{E_h}+\ket{0}_{B_h}\ket{\pm2}_{E_h}}{\sqrt{2}}.
 \label{eq:qubit-erasure}
\end{equation}
This is a qubit half-erasure channel, with transmitted basis states
$\ket{\pm2}$ and erasure flag $\ket{0}$. The environment $\ket{\xi}$
is magic: its computational-basis support has size two, whereas a
pure single-qudit stabilizer state in prime dimension has support
of size $1$ or $d$~\cite{Gross2006}.

\begin{theorem}[Superactivation of two convolution channels]
\label{thm:two-convolutions}
Let $\N_{G,\sigma}$ be any of the channels constructed in
Theorem~\ref{thm:private}, and let $\mathcal A_d$ be the convolution
auxiliary defined above. Then
\[
 Q(\N_{G,\sigma})=Q(\mathcal A_d)=0,~
 Q(\N_{G,\sigma}\otimes\mathcal A_d)>0.
\]
With both interactions fixed, replacing either environment alone
by any stabilizer mixture on that register reduces the joint
quantum capacity to zero.
\end{theorem}

\begin{proof}[Proof sketch]
Alice encodes the five qubits of $C$ into a data block and protects the encoding label
by repetition. Each data or label qubit uses the half-erasure
subspace of $\mathcal A_d$. From the surviving label and erasure
positions, Bob either recovers $C$ exactly, with all its reference
correlations, or outputs an erasure flag. The failure probability
$p$ is input independent. For finite $L$, the isometric encoder
$\mathcal V_d$ and decoder $\mathcal D_d$ of Sec.~SIV of the
Supplemental Material~\cite{supplement} give
$1/2<p<33823/65536<p_*$ and satisfy
\begin{equation}
 \mathcal D_d\circ\mathcal A_d^{\otimes L}\circ\mathcal V_d
 =\mathcal E_p^{(32)}.
 \label{eq:encoded-erasure}
\end{equation}

For $\omega^G_{AC}$ from Theorem~\ref{thm:private}, Alice sends
$A$ through $\N_{G,\sigma}$ and encodes only $C$.
Equation~\eqref{eq:encoded-erasure} gives coherent information
$f(p)>0$ after decoding; by data processing it is at least as
large before decoding. This uses the main channel once and the
auxiliary $L$ times. Independent pure states on the other $L-1$ main
inputs add zero coherent information, giving $L$ uses of the pair.
Hence
\begin{equation}
 Q(\N_{G,\sigma}\otimes\mathcal A_d)\ge \frac{f(p)}{L}>0.
 \label{eq:two-convolution-rate}
\end{equation}

Replacing either environment by a stabilizer mixture makes the
corresponding channel entanglement breaking~\cite{Xiong2026}.
Such a channel has a measure-and-prepare representation and a
separable Choi state~\cite{Horodecki2003}. If the main environment
is replaced, that channel and the self-complementary auxiliary
are both antidegradable, so their product is antidegradable.
If the auxiliary environment is replaced, its Choi state is PPT,
as is its product with the PPT Choi state of the main channel.
In either case the joint quantum capacity is zero.
\end{proof}

The state of the first environment also limits the rate of joint
quantum communication. For every positive BGJ coupling $G$,
every environment state $\sigma$, and every finite-dimensional
antidegradable auxiliary $\mathcal A$, we obtain~\cite{supplement}
\[
 Q(\mathcal N_{G,\sigma}\otimes\mathcal A)
 \le \operatorname{MRM}(\sigma).
\]
The modified relative entropy of magic $\operatorname{MRM}(\sigma)$
is the minimum of the quantum relative entropy $D_{\rm rel}(\sigma\Vert\tau)$ over normalized
projectors $\tau$ onto joint eigenspaces of commuting Weyl
operators~\cite{BGJ2026,XiongEtAl2026Transport}. This covers one main-channel use
with the antidegradable auxiliary $\mathcal A_d^{\otimes L}$.

For the $(7,7)$ example, superactivation persists on an open set
of main-channel environment states, with both interactions and the
auxiliary environment fixed~\cite{supplement}. Parentin \emph{et al.}
showed that finitely many joint uses of a PPT channel and a
half-erasure channel can transmit a qubit at a fidelity unattainable
with either channel alone~\cite{Parentin2026}.
It remains open how the choice of magic environments affects the
number of uses needed for such an advantage with two convolution channels.

\paragraph*{Acknowledgments.}
We thank Hongyin Zhao, Qinghua Zhang, and Changliang Ren for helpful
discussions and Marco Parentin for drawing our attention to
Ref.~\cite{Parentin2026}. This work was supported by the National Natural Science
Foundation of China (Grant Nos. 12671165 and 12201555), the Natural Science Foundation
of Hunan Province (Grant No. 2025JJ50050), and the Hunan Basic Science
Research Center for Mathematical Analysis (Grant No. 2024JC2002).

\paragraph*{Data Availability.}
The article and its Supplemental Material give the analytical
constructions, proofs, and equations for the plotted curves.
The plotting and verification scripts accompany the Supplemental Material.

\paragraph*{Conflicts of Interest.}
The authors declare no conflicts of interest.

\paragraph*{AI Use Disclosure.}
The authors used OpenAI ChatGPT (GPT-6.0 Astra and GPT-6.2 Astra)
to help develop the constructions, check algebraic steps, prepare
verification and plotting scripts, and draft and revise the text.
The authors specified the channel model and directed the derivation of
the private-information witness and the auxiliary code. They verified
the derivations and calculations by hand and take responsibility for
the proofs and the final manuscript.

\bibliographystyle{apsrev4-2}
\bibliography{references}
\fi

\ifPrintSupplement
\ifPrintLetter
\clearpage
\onecolumngrid
\fi

\setcounter{equation}{0}
\setcounter{section}{0}
\setcounter{subsection}{0}
\setcounter{figure}{0}
\setcounter{table}{0}
\setcounter{theorem}{0}
\setcounter{lemma}{0}
\setcounter{proposition}{0}
\setcounter{corollary}{0}
\setcounter{secnumdepth}{2}
\renewcommand{\theequation}{S\arabic{equation}}
\renewcommand{\thesection}{S\Roman{section}}
\renewcommand{\thetable}{S\arabic{table}}
\renewcommand{\thefigure}{S\arabic{figure}}
\renewcommand{\thetheorem}{S\arabic{theorem}}
\renewcommand{\thelemma}{S\arabic{lemma}}
\renewcommand{\theproposition}{S\arabic{proposition}}
\renewcommand{\thecorollary}{S\arabic{corollary}}
\renewcommand{\theHequation}{supp.\arabic{equation}}
\renewcommand{\theHsection}{supp.\arabic{section}}
\renewcommand{\theHsubsection}{supp.\arabic{section}.\arabic{subsection}}
\renewcommand{\theHfigure}{supp.\arabic{figure}}
\renewcommand{\theHtable}{supp.\arabic{table}}
\renewcommand{\theHtheorem}{supp.\arabic{theorem}}
\renewcommand{\theHlemma}{supp.\arabic{lemma}}
\renewcommand{\theHproposition}{supp.\arabic{proposition}}
\renewcommand{\theHcorollary}{supp.\arabic{corollary}}

\ifPrintLetter
\begin{center}
{\large\bfseries Supplemental Material for\\[3pt]
Magic Can Superactivate Quantum Capacity}\par\medskip
Chunhe Xiong, Sunho Kim, Shunlong Luo, Lin Zhang, Shao-Ming Fei, and Junde Wu\par
\smallskip
{\small Author affiliations and contact information are given in the Letter.}
\end{center}
\medskip
\else
\title{Supplemental Material for\texorpdfstring{\\}{ }Magic Can Superactivate Quantum Capacity}
\PaperAuthors
\maketitle
\fi

\section{Conventions and information quantities}
\label{sm:sec:operational-model}

Let $d$ be an odd prime, $V=\F_d^n$, and $D=d^n$. Each of the
registers $A,Y,B,E$ consists of $n$ qudits. For $x,y\in V$,
write $\ket{x,y}_{AY}=\ket{x}_A\otimes\ket{y}_Y$, and likewise
for $BE$. Following the convolution convention of
Refs.~\cite{BGJ2023,BGJ2026}, write
$G^{-\mathsf T}=(G^{-1})^{\mathsf T}$ for $G\in\mathrm{GL}(2,\F_d)$
and set
\begin{equation}
 \begin{aligned}
 G^{-\mathsf T}
   &=\begin{pmatrix}a&b\\c&e\end{pmatrix},\\
 U_G\ket{x,y}_{AY}
   &=\ket{ax+by,cx+ey}_{BE},\\
 \N_{G,\sigma}(\rho)
   &=\Tr_E\!\left[U_G(\rho_A\otimes\sigma_Y)U_G^\dagger\right].
 \end{aligned}
 \label{sm:eq:model-convolution}
\end{equation}
Here $A$ is the signal input, $Y$ the environment input, and $B,E$
the corresponding outputs.
The matrix $G^{-\mathsf T}$ acts on each coordinate pair
$(x_k,y_k)$ over $\F_d$. Since this map is invertible, $U_G$
permutes the computational basis and is therefore unitary.
A coupling is called positive if all
four entries of $G$, or equivalently of $G^{-\mathsf T}$, are nonzero.

The environment state $\sigma_Y$ is fixed and independent of the
signal input. Choose a purification $\ket{\Sigma}_{YF}$ and define
the Stinespring isometry $V_\sigma:A\to BK$, where $K=EF$, by
\begin{equation}
 \begin{aligned}
 V_\sigma\ket{\psi}_A
   &=(U_G\otimes I_F)
     (\ket{\psi}_A\otimes\ket{\Sigma}_{YF}),\\
 \N_{G,\sigma}(\rho)
   &=\Tr_K\!\left[V_\sigma\rho V_\sigma^\dagger\right],\qquad
 \N_{G,\sigma}^c(\rho)
   =\Tr_B\!\left[V_\sigma\rho V_\sigma^\dagger\right].
 \end{aligned}
 \label{sm:eq:model-complement}
\end{equation}
Bob receives $B$; Eve receives the full complementary output
$K=EF$, including the environment purifier. For $\ell$ uses the
initial environment state is $\sigma_Y^{\otimes\ell}$. Alice may
encode jointly across all channel inputs, and Bob may decode
jointly across all his outputs.

Write $\N=\N_{G,\sigma}$, and denote its unassisted private
classical capacity and quantum capacity by $P(\N)$ and $Q(\N)$,
respectively.
To compare Bob's and Eve's information about the same signal label,
suppose that Alice draws $t$ with probability $p_t$, records it in
a classical register $T$, and sends $\rho_t$ through $\N$.
The joint state of her record and the two outputs is then
\begin{equation}
 \Omega_{TBK}
 =\sum_t p_t\proj{t}_T\otimes V_\sigma\rho_t V_\sigma^\dagger,
 \qquad
 B_t=\N(\rho_t),\qquad K_t=\N^c(\rho_t).
 \label{sm:eq:model-cq-state}
\end{equation}
Here $B_t$ and $K_t$ are the states received by Bob and Eve,
respectively, conditional on Alice's choice $t$.

For either output $O=B,K$, write $\bar O=\sum_t p_t O_t$.
Since the marginal $\Omega_{TO}$ is block diagonal in the label $t$,
its von Neumann entropy is
$S(\Omega_{TO})=H(T)+\sum_t p_t S(O_t)$, where $H(T)$ is the
Shannon entropy of $\{p_t\}$. All entropies use base-two
logarithms. Consequently,
\begin{equation}
 \begin{aligned}
 I(T;O)_\Omega
   &= S(\bar O)-\sum_t p_t S(O_t),\\
 H(T|O)_\Omega
   &:= S(\Omega_{TO})-S(\bar O)
    = H(T)-I(T;O)_\Omega .
 \end{aligned}
 \label{sm:eq:model-holevo}
\end{equation}
The first line is the Holevo information of the output ensemble.
The private information of the chosen input ensemble is
$\Delta:=I(T;B)_\Omega-I(T;K)_\Omega$. Both terms refer to
the same label, so $H(T)$ cancels when their difference is written
in terms of conditional entropies. The private coding
theorem~\cite{Devetak2005} then gives
\begin{equation}
 P(\N)\ge \Delta
 =H(T|K)_\Omega-H(T|B)_\Omega .
 \label{sm:eq:model-private-bound}
\end{equation}
This identity for $\Delta$ holds for arbitrary quantum outputs.
When the conditional states $O_t$ are diagonal in a common basis,
$H(T|O)_\Omega$ can also be computed as the classical conditional
entropy of $T$ given the outcome in that basis. We use this fact
for each output in the seven-qudit example below.

For quantum communication, the coherent information of an input
state $\rho$ and the corresponding coding bound are
\begin{equation}
 I_c(\rho,\N)=S(\N(\rho))-S(\N^c(\rho)),\qquad
 Q(\N)\ge \frac{1}{\ell}I_c(\rho_\ell,\N^{\otimes\ell}),
 \label{sm:eq:model-coherent-information}
\end{equation}
where $\rho_\ell$ is any joint input state for $\ell$ uses.
Both capacity bounds follow from the regularized coding
formulas~\cite{Devetak2005}:
\[
 X(\N)=\sup_{\ell\ge1}
       \frac{X^{(1)}(\N^{\otimes\ell})}{\ell},
 \qquad X=P,Q,
\]
where $P^{(1)}$ is the supremum of $\Delta$ over finite input
ensembles and $Q^{(1)}$ is the maximum of $I_c$ over input states.

For the stabilizer conventions, let $X\ket j=\ket{j+1}$ and
$Z\ket j=\omega_d^j\ket j$, where $j\in\F_d$ and
$\omega_d=e^{2\pi i/d}$. For $z=(u,v)$ and $z'=(u',v')$ in
$V\oplus V$, set
\[
 W(z)=\omega_d^{-2^{-1}u\cdot v}
 \bigotimes_{k=1}^n Z^{u_k}X^{v_k},\qquad
 [z,z']=u\cdot v'-v\cdot u'.
\]
Here $2^{-1}$ is taken in $\F_d$. The relation
$W(z)W(z')=\omega_d^{[z,z']}W(z')W(z)$ shows that Weyl
operators commute on any Lagrangian subspace $M$, that is, an
$n$-dimensional subspace on which the symplectic form vanishes.
For a linear functional $\ell:M\to\F_d$, the projector onto the
common eigenspace with eigenvalues $\omega_d^{\ell(z)}$ is
\[
 \Pi_{M,\ell}=\frac1D\sum_{z\in M}\omega_d^{-\ell(z)}W(z).
\]
These rank-one projectors are the pure $n$-qudit stabilizer states.
Their convex hull $\STAB_n$ includes mixtures of entangled as well
as product stabilizer states. States outside $\STAB_n$ are called
magic~\cite{Gross2006,Veitch2014}. Clifford unitaries preserve the
Weyl operators up to phase under conjugation.

Section~\ref{sm:sec:example} proves the $(7,7)$ example and its
activation by a half-erasure channel. Section~\ref{sm:sec:theorem-one}
extends the construction to the couplings of Theorem~1.
Section~\ref{sm:sec:theorem-two} constructs the convolution auxiliary
code, proves Theorem~2, and derives the environmental capacity bound
and robustness results.

\section{A seven-qudit private channel and its two auxiliaries}
\label{sm:sec:example}\label{sm:sec:simple-example}

\subsection{The construction}
Set $d=n=7$, $V=\F_7^7$, and $D=7^7$. Define
\begin{equation}
 U\ket{x,y}_{AY}=\ket{x+y,y-2x}_{BE},\qquad
 G_{\mathrm{priv}}=3^{-1}\begin{pmatrix}1&2\\-1&1\end{pmatrix}
 =\begin{pmatrix}5&3\\2&5\end{pmatrix}.
 \label{sm:eq:example-coupling}
\end{equation}
Thus $G_{\mathrm{priv}}^{-\mathsf T}=
\left(\begin{smallmatrix}1&1\\-2&1\end{smallmatrix}\right)$.
For each $(u,v)\in\{0,1,2,3\}^2$, define
\begin{equation}
 a_{4u+v}=(u,v,u^2-3v^2)\in\F_7^3.
 \label{sm:eq:example-points}
\end{equation}
This gives sixteen distinct vectors, each with three coordinates. The
index $4u+v$ runs through the integers $0,\ldots,15$; only the vector
coordinates are evaluated modulo $7$.
The labels $r_i,s_j$ specify the signal states on $A$, while
$u_{ij},v_{ij},w_i,z_j$ specify the environment state on $Y$.
With $0_3=(0,0,0)$, define
\begin{equation}
 \begin{aligned}
 r_i&=(1,a_i,0_3),&s_j&=(0,0_3,a_j),\\
 u_{ij}&=-r_i-3s_j,&v_{ij}&=-3r_i-s_j,\\
 w_i&=-4r_i,&z_j&=-4s_j,
 \end{aligned}
 \qquad 0\le i,j\le15.
 \label{sm:eq:simple-labels}
\end{equation}
Let $m=16$ be the number of basis states in each signal mixture,
set $Z=640$, and define the real Hadamard matrix
$H_{16}=\left(\begin{smallmatrix}1&1\\1&-1\end{smallmatrix}\right)^{\otimes4}$.
Write $H_{ij}=(H_{16})_{ij}$, so that
$H_{16}H_{16}^{\mathsf T}=16I_{16}$.
The environment and
binary input ensemble are
\begin{equation}
 \begin{aligned}
 \ket{\phi_{ij}}&=\ket{u_{ij}}+H_{ij}\ket{v_{ij}},~
 \sigma=\frac1Z\left[
       \sum_{i,j}\proj{\phi_{ij}}
       +4\sum_i\proj{w_i}+4\sum_j\proj{z_j}\right],\\
 \rho_0&=\frac1m\sum_i\proj{r_i},~
 \rho_1=\frac1m\sum_j\proj{s_j},~\Pr(T=0)=\Pr(T=1)=\tfrac12.
 \end{aligned}
 \label{sm:eq:simple-state}
\end{equation}
The vectors $\ket{\phi_{ij}}$ are unnormalized. The signs $H_{ij}$,
the weights $4$, and the normalization $Z$ are real numbers.
The first coordinates of $u_{ij},v_{ij},w_i,z_j$ are $6,4,3,0$,
respectively, and the remaining coordinates distinguish the labels
within each family. Thus the environment state has nonzero diagonal
entries at exactly $2m^2+2m=544$ computational-basis labels.

On each pair $\{\ket{u_{ij}},\ket{v_{ij}}\}$, the corresponding
block of $Z\sigma$ is
$\left(\begin{smallmatrix}1&H_{ij}\\ H_{ij}&1\end{smallmatrix}\right)$,
with eigenvalues $2$ and $0$. Each of the $2m$ remaining labels
contributes a one-dimensional block of weight $4$. Hence
\[
 \Tr\sigma=\frac{2m^2+8m}{Z}=1,\qquad
 \rank\sigma=m^2+2m=288.
\]
Write $\N_\sigma:=\N_{G_{\mathrm{priv}},\sigma}$ for the resulting
channel.

\begin{proposition}[A PPT channel with positive private capacity]
\label{sm:prop:example-private}
Let $J$ be the normalized Choi state of $\N_\sigma$, and let
$T_B$ denote partial transpose on Bob's output. Then
\begin{equation}
 J^{T_B}\succeq0,\qquad
 \rank J=288D,\qquad \rank J^{T_B}=528D.
 \label{sm:eq:simple-choi-ranks}
\end{equation}
The ensemble in Eq.~\eqref{sm:eq:simple-state} has private
information $\Delta=\epsrate:=(12/5-\log_2 5)/32>0$.
Consequently, $Q(\N_\sigma)=0$ and $P(\N_\sigma)\ge\epsrate$.
\end{proposition}

\subsection{Proof of the private PPT witness}
\label{sm:sec:simple-ppt-certificate}

\begin{proof}
\emph{The PPT certificate.}
Let $R_A\simeq A$ be the reference register, and write
$\sigma_{y,y'}=\bra{y}\sigma\ket{y'}$. In the following Choi
matrices, the first basis label belongs to $R_A$ and the second
to $B$. For the input operator $\ket{x}\bra{x'}$, tracing out
$E$ imposes $y-2x=y'-2x'=:e$. The normalized Choi state is
therefore
\begin{equation}
 \begin{aligned}
 J
 &=\frac1D\sum_{x,x'\in V}
      \ket{x}\bra{x'}\otimes\N_\sigma(\ket{x}\bra{x'})\\
 &=\frac1D\sum_{x,x',y,y'\in V}
      \sigma_{y,y'}\,\delta_{y-2x,y'-2x'}\,
      \ket{x,x+y}\bra{x',x'+y'}\\
 &=\frac1D\sum_{e,x,x'\in V}
      \sigma_{e+2x,e+2x'}\,
      \ket{x,e+3x}\bra{x',e+3x'} .
 \end{aligned}
 \label{sm:eq:example-choi-expansion}
\end{equation}
In the last line, the Bob coordinate minus three times the
reference coordinate equals $e$ on both sides of each matrix
element. Different values of $e$ thus label orthogonal invariant
subspaces. The orthonormal basis
\[
 \ket{k_{e,u}}
 =\ket{(u-e)/2,(3u-e)/2},
 \qquad e,u\in V,
\]
satisfies
\[
 \bra{k_{e,u}}J\ket{k_{e',v}}
 =\frac{\delta_{e,e'}}D\,\sigma_{u,v}.
\]

Partial transpose on $B$ gives
\[
 J^{T_B}
 =\frac1D\sum_{e,x,x'\in V}
    \sigma_{e+2x,e+2x'}\,
    \ket{x,e+3x'}\bra{x',e+3x}.
\]
Now the Bob coordinate plus three times the reference coordinate
has the same value on both sides, namely $s=e+3(x+x')$.
Use the orthonormal basis
\[
 \ket{h_{s,u}}
 =\ket{(s-u)/4,(s+3u)/4},
 \qquad s,u\in V.
\]
Within the same $s$ block, substituting $x=(s-u)/4$ and
$x'=(s-v)/4$ into $e=s-3(x+x')$ gives
\[
 e+2x=(u+3v)/4,\qquad e+2x'=(3u+v)/4.
\]
Hence
\[
 \bra{h_{s,u}}J^{T_B}\ket{h_{s',v}}
 =\frac{\delta_{s,s'}}D\,
   \sigma_{(u+3v)/4,(3u+v)/4}.
\]
Defining
\[
 [\mathcal R_3(X)]_{u,v}
 :=X_{(u+3v)/4,(3u+v)/4},
\]
we obtain
\begin{equation}
 J\simeq\bigoplus_{e\in V}\frac{\sigma}{D},
 \qquad
 J^{T_B}\simeq\bigoplus_{s\in V}
                  \frac{\mathcal R_3(\sigma)}{D}.
 \label{sm:eq:example-choi-blocks}
\end{equation}
Here $\simeq$ denotes a simultaneous permutation of rows and
columns. Each direct sum contains $D$ blocks of size $D$.
The map $\mathcal R_3$, by contrast, rearranges the entries
within an environment matrix and can change its rank.
It fixes the diagonal matrix units and acts on an arbitrary
matrix unit as
\[
 \mathcal R_3(\ket{y}\bra{z})
 =\ket{(3z-y)/2}\bra{(3y-z)/2}.
\]
Since
\[
 (3v_{ij}-u_{ij})/2=w_i,\qquad
 (3u_{ij}-v_{ij})/2=z_j,
\]
the matrix unit $\ket{u_{ij}}\bra{v_{ij}}$ moves to
$\ket{w_i}\bra{z_j}$, and its adjoint moves to
$\ket{z_j}\bra{w_i}$. The diagonal entries at $u_{ij}$ and
$v_{ij}$ remain in place. Thus
\begin{equation}
 Z\mathcal R_3(\sigma)
 \simeq I_{512}\oplus
 \begin{pmatrix}
   4I_{16}&H_{16}\\
   H_{16}^{\mathsf T}&4I_{16}
 \end{pmatrix}
 \oplus 0 .
 \label{sm:eq:simple-pt-block}
\end{equation}
Using $H_{16}^{\mathsf T}H_{16}=16I_{16}$, the middle block
factors as
\[
 \begin{pmatrix}
   2I_{16}\\ H_{16}^{\mathsf T}/2
 \end{pmatrix}
 \begin{pmatrix}
   2I_{16}&H_{16}/2
 \end{pmatrix}.
\]
It is positive semidefinite and has rank $16$, since the first
factor has full column rank. Equation~\eqref{sm:eq:example-choi-blocks}
and $\rank\sigma=288$ now give
\[
 J^{T_B}\succeq0,\qquad
 \rank J=288D,\qquad
 \rank J^{T_B}=(512+16)D=528D.
\]
A channel with a PPT Choi state has zero quantum capacity;
hence $Q(\N_\sigma)=0$~\cite{Horodecki1998,Horodecki2000Binding}.

\emph{Outputs of the private ensemble.}
Choose the environment purification
\begin{equation}
 \ket{\Sigma}_{YF}
 =\frac1{\sqrt Z}\left[
   \sum_{i,j}
    \bigl(\ket{u_{ij}}+H_{ij}\ket{v_{ij}}\bigr)\ket{ij}_F
   +2\sum_i\ket{w_i}\ket{w,i}_F
   +2\sum_j\ket{z_j}\ket{z,j}_F
 \right],
 \label{sm:eq:simple-purification}
\end{equation}
where the $288$ flags are orthonormal and $ij$ denotes the
ordered pair $(i,j)$. The two endpoints $u_{ij},v_{ij}$ share
one flag, so tracing out $F$ retains their off-diagonal terms
and gives exactly the state $\sigma$ in Eq.~\eqref{sm:eq:simple-state}.

Let $\mathcal Y$ be the set of $544$ environment basis labels
in Eq.~\eqref{sm:eq:simple-purification}. Denote their flags and unnormalized amplitudes by
\[
 \begin{aligned}
 f(u_{ij})&=f(v_{ij})=ij,&
 f(w_i)&=(w,i),& f(z_j)&=(z,j),\\
 c_{u_{ij}}&=1,& c_{v_{ij}}&=H_{ij},&
 c_{w_i}&=c_{z_j}=2 .
 \end{aligned}
\]
Put $q_y=|c_y|^2$. For a computational-basis input $\ket{x}_A$,
the isometry of Eq.~\eqref{sm:eq:model-complement} gives
\begin{equation}
 \ket{\Psi_x}_{BEF}
 =V_\sigma\ket{x}_A
 =\frac1{\sqrt Z}\sum_{y\in\mathcal Y}
   c_y\ket{x+y}_B\ket{y-2x}_E\ket{f(y)}_F .
 \label{sm:eq:simple-pure-output}
\end{equation}
For fixed $x$,
\[
 \begin{aligned}
 \langle y'-2x,f(y')\mid y-2x,f(y)\rangle
   =\delta_{y,y'},~
 \langle x+y'\mid x+y\rangle
   =\delta_{y,y'} .
 \end{aligned}
\]
The first equality holds even for two endpoints with the same
flag, since $y\mapsto y-2x$ is injective. Tracing out $EF$ or
$B$, respectively, therefore gives
\begin{equation}
 \begin{aligned}
 B(x)&=\frac1Z\sum_{y\in\mathcal Y}q_y\proj{x+y},\\
 K(x)&=\frac1Z\sum_{y\in\mathcal Y}q_y\proj{y-2x,f(y)},
 \qquad K=EF .
 \end{aligned}
 \label{sm:eq:simple-marginals}
\end{equation}
By linearity, the conditional outputs of the two signal
mixtures are
\[
 \begin{aligned}
 B_0&=\frac1m\sum_\kappa B(r_\kappa),&
 B_1&=\frac1m\sum_\ell B(s_\ell),\\
 K_0&=\frac1m\sum_\kappa K(r_\kappa),&
 K_1&=\frac1m\sum_\ell K(s_\ell).
 \end{aligned}
\]
The classical register $T$ records which mixture was prepared,
not its component $\kappa$ or $\ell$. The pair flag $ij$ in $F$
labels a term in the environment purification. By
Eq.~\eqref{sm:eq:simple-marginals}, $B_0,B_1$ are diagonal in
the computational basis of $B$, and $K_0,K_1$ are diagonal in
the computational and flag basis of $EF$.

\emph{Positive private information.}
For the equiprobable ensemble in Eq.~\eqref{sm:eq:simple-state},
Eq.~\eqref{sm:eq:model-private-bound} gives
\[
 P(\N_\sigma)\ge\Delta=H(T|K)-H(T|B).
\]
For either output $O=B,K$, let $o$ denote an outcome in its
common diagonal basis, and write
\[
 a_o=\Pr(T=0,O=o)=\tfrac12\bra{o}O_0\ket{o},\qquad
 b_o=\Pr(T=1,O=o)=\tfrac12\bra{o}O_1\ket{o}.
\]
Then
\[
 \Omega_{TO}=\sum_o\bigl(a_o\proj{0,o}+b_o\proj{1,o}\bigr),
 \qquad
 \Omega_O=\sum_o(a_o+b_o)\proj{o}.
\]
Taking the difference of their entropies gives
\begin{equation}
 \begin{aligned}
 H(T|O)
 &=-\sum_o\bigl(a_o\log_2a_o+b_o\log_2b_o\bigr)
   +\sum_o(a_o+b_o)\log_2(a_o+b_o)\\
 &=\sum_{o:\,a_o+b_o>0}
       (a_o+b_o)\,
       \hb\!\left(\frac{a_o}{a_o+b_o}\right),
 \end{aligned}
 \label{sm:eq:conditional-entropy-overlap}
\end{equation}
where $\hb(p)=-p\log_2p-(1-p)\log_2(1-p)$ and $0\log_2 0=0$.
If $o$ can arise from only one signal, then
$\Pr(T=0\mid O=o)$ is $0$ or $1$, and this outcome contributes
zero to $H(T|O)$. It therefore suffices to count the
\emph{ambiguous outputs}, those with $a_o,b_o>0$.

A signal component $x$ and an environment label $y$ contribute
the joint probability
\[
 \frac12\cdot\frac1m\cdot\frac{q_y}{Z}
 =\frac{q_y}{2mZ}.
\]
Here $q_{u_{ij}}=q_{v_{ij}}=1$ and $q_{w_i}=q_{z_j}=4$.
Contributions with the same output label are added.

\emph{Bob's outputs.}
The table gives the first coordinate of Bob's output $x+y$
in $\F_7$. Its rows specify the signal and its columns the
family of the environment label $y$:
\[
 \begin{array}{c|rrrr}
       &u_{ij}&v_{ij}&w_i&z_j\\ \hline
  T=0  &0&-2&-3&1\\
  T=1  &-1&-3&-4&0
 \end{array}
\]
Only $0$ and $-3$ occur in both rows. At first coordinate $0$,
a common output must satisfy
$r_\kappa+u_{ij}=s_\ell+z_{j'}$. Comparing the remaining
coordinate blocks gives
\[
 a_\kappa=a_i,\qquad a_\ell+3a_j=4a_{j'} .
\]
For the points in Eq.~\eqref{sm:eq:example-points},
\begin{equation}
 a_i+3a_j=4a_\ell\quad\Longrightarrow\quad i=j=\ell .
 \label{sm:eq:simple-code}
\end{equation}
To verify this, write
$a_i=(\xi,q(\xi))$, $a_j=(\eta,q(\eta))$, and
$a_\ell=(\zeta,q(\zeta))$, where
$q:\F_7^2\to\F_7$ is given by
$q(\xi)=\xi_1^2-3\xi_2^2$.
The first two coordinates give $\zeta=(\xi+3\eta)/4$;
the last coordinate then gives
\[
 0=q(\xi)+3q(\eta)-4q\!\left(\frac{\xi+3\eta}{4}\right)
   =\frac34q(\xi-\eta).
\]
The nonzero squares in $\F_7$ are $1,2,4$.
Since $3$ is a nonsquare, $q(\delta)=0$ implies $\delta=0$.
Thus $\xi=\eta=\zeta$, proving Eq.~\eqref{sm:eq:simple-code}.

It follows that $\kappa=i$ and $\ell=j=j'$. The common output is
$-3s_j$. For fixed $j$, signal $0$ contributes the $m$ terms
$r_i+u_{ij}=-3s_j$, each of weight $1/(2mZ)$.
Signal $1$ contributes the single term $s_j+z_j=-3s_j$
of weight $4/(2mZ)$.

At first coordinate $-3$, a common output must satisfy
$r_\kappa+w_i=s_\ell+v_{i'j}$. Its remaining coordinate blocks
give
\[
 a_\kappa+3a_{i'}=4a_i,\qquad a_\ell=a_j .
\]
Equation~\eqref{sm:eq:simple-code} now gives $\kappa=i'=i$
and $\ell=j$. The common output is $-3r_i$.
For fixed $i$, signal $0$ contributes $r_i+w_i=-3r_i$,
whereas signal $1$ contributes the $m$ terms
$s_j+v_{ij}=-3r_i$. Thus Bob's complete list of ambiguous
outputs and their joint probabilities is
\[
 \begin{array}{c|cc}
 b & \Pr(T=0,B=b) & \Pr(T=1,B=b)\\ \hline
 -3s_j & m/(2mZ) & 4/(2mZ)\\
 -3r_i & 4/(2mZ) & m/(2mZ)
 \end{array}
\]
There are $m$ outputs of each type. Each has probability
$(m+4)/(2mZ)$, and its posterior probability for $T=0$ is
$m/(m+4)=4/5$ or $4/(m+4)=1/5$, respectively.

\emph{Eve's outputs.}
Eve receives $K=EF$, so a common output requires equality
of both the $E$ coordinate and the flag in $F$. With a
singleton flag, or with the same endpoint of a pair flag,
equality of the $E$ coordinates would require
\[
 y-2r_\kappa=y-2s_\ell .
\]
This is impossible because $r_\kappa$ and $s_\ell$ have
different first coordinates.

It remains to compare opposite endpoints with the same flag
$ij$. In one orientation,
\[
 \begin{aligned}
 u_{ij}-2r_\kappa=v_{ij}-2s_\ell
 &\quad\Longleftrightarrow\quad
 r_i-r_\kappa=s_j-s_\ell\\
 &\quad\Longleftrightarrow\quad
 \kappa=i,\quad\ell=j .
 \end{aligned}
\]
Indeed, the two differences are supported on disjoint
coordinate blocks, so equality forces both to vanish.
The common output is $(-3r_i-3s_j,ij)$.
The reverse orientation would require
\[
 v_{ij}-2r_\kappa=u_{ij}-2s_\ell
 \quad\Longleftrightarrow\quad
 r_i+r_\kappa=s_j+s_\ell ,
\]
whose first coordinates would give $2=0$ in $\F_7$.
Eve's ambiguous outputs and their joint probabilities are
therefore exactly
\[
 \begin{array}{c|cc}
 k & \Pr(T=0,K=k) & \Pr(T=1,K=k)\\ \hline
 (-3r_i-3s_j,ij) & 1/(2mZ) & 1/(2mZ)
 \end{array}
\]
The distinct pair flags give $m^2$ such outputs. Each has
probability $1/(mZ)$ and equal posterior probabilities for
the two signals.

Substituting the two tables into
Eq.~\eqref{sm:eq:conditional-entropy-overlap}, with
$m=16$ and $Z=640$, gives
\begin{equation}
 \begin{aligned}
 H(T|B)
 &=2m\,\frac{m+4}{2mZ}\,
      \hb\!\left(\frac4{m+4}\right)
   =\frac{\hb(1/5)}{32},\\
 H(T|K)
 &=m^2\,\frac{2}{2mZ}\,\hb(1/2)
   =\frac1{40}.
 \end{aligned}
 \label{sm:eq:simple-conditional-entropies}
\end{equation}
Since $H(T)=1$, the corresponding Holevo informations are
\begin{equation}
 I(T;B)=1-\frac{\hb(1/5)}{32},\qquad
 I(T;K)=\frac{39}{40}.
 \label{sm:eq:simple-informations}
\end{equation}
Their difference is
\begin{equation}
 \Delta
 =H(T|K)-H(T|B)
 =\frac1{40}-\frac{\hb(1/5)}{32}
 =\frac{12/5-\log_2 5}{32}
 =\epsrate>0 .
 \label{sm:eq:simple-leakage}
\end{equation}
Here $\hb(1/5)=\log_2 5-8/5$, and strict positivity follows
from $5^5<2^{12}$. Equation~\eqref{sm:eq:model-private-bound}
therefore gives $P(\N_\sigma)\ge\epsrate$.
\end{proof}

The weight $4$ has two roles in this construction. In the middle
block of Eq.~\eqref{sm:eq:simple-pt-block}, replacing both diagonal
blocks by $cI_{16}$ gives a smallest eigenvalue of $c-4$, since
all singular values of $H_{16}$ are $4$. Thus $4$ is the smallest
common diagonal weight for which this block is positive semidefinite.

For the chosen signal ensemble, the Hadamard signs enter the
output probabilities only through $|H_{ij}|^2=1$. At Bob's output
$-3s_j$, the $16$ unit-weight contributions $r_i+u_{ij}$ from
signal $0$ give the same label as the single weight-$4$
contribution $s_j+z_j$ from signal $1$. At Eve's output
$(-3r_i-3s_j,ij)$, the two endpoints of the same coherent pair
share the flag $ij$ and contribute one unit of weight from each
signal.

\subsection{Smith--Yard activation and the erasure window}
\label{sm:sec:erasure-activation}\label{sm:sec:smerasure}

Each signal in Eq.~\eqref{sm:eq:simple-state} has rank $16$.
Let $C=C_0\oplus C_1$, where each $C_t$ has orthonormal basis
$\{\ket{t,i}:0\le i\le15\}$. Choose the conditional purifications
\begin{equation}
 \begin{aligned}
 \ket{\psi_0}_{AC}&=\frac14\sum_i\ket{r_i}_A\ket{0,i}_C,\\
 \ket{\psi_1}_{AC}&=\frac14\sum_j\ket{s_j}_A\ket{1,j}_C .
 \end{aligned}
 \label{sm:eq:smpsit}
\end{equation}
Thus $\Tr_C\proj{\psi_t}=\rho_t$, and the two purifications
occupy orthogonal sectors of $C$, giving $\dim C=16+16=32$.
The input to the joint channel is the mixture
\begin{equation}
 \omega_{AC}=\tfrac12\proj{\psi_0}+\tfrac12\proj{\psi_1}.
 \label{sm:eq:smomega}
\end{equation}
Alice sends $A$ through $\N_\sigma$ and $C$ through the erasure
channel. A separate two-dimensional reference $R_0$ purifies
the entire mixture:
\[
 \ket{\Psi}_{R_0AC}
 =\frac{\ket{0}_{R_0}\ket{\psi_0}_{AC}
       +\ket{1}_{R_0}\ket{\psi_1}_{AC}}{\sqrt2}.
\]
The reference $R_0$ is untouched. The register $F$ in
Eq.~\eqref{sm:eq:simple-purification}, by contrast, purifies
the environment of $\N_\sigma$ and belongs to its complementary
output $K=EF$.

For $0\le p\le1$, the erasure channel on $C$ is
\[
 \E_p^{(32)}(X)=(1-p)X\oplus p\Tr(X)\proj{\mathrm{er}},
\]
where the unit vector $\ket{\mathrm{er}}$ is orthogonal to the
$32$-dimensional data subspace. Identifying the data subspaces of
the receiver's output $B'$ and the complementary output $E'$ with
$C$, a Stinespring isometry for this channel is
\[
 W_p\ket{\phi}_C
 =\sqrt{1-p}\,\ket{\phi}_{B'}\ket{\mathrm{er}}_{E'}
  +\sqrt p\,\ket{\mathrm{er}}_{B'}\ket{\phi}_{E'}.
\]
Orthogonality of the data subspace and the flag gives
$W_p^\dagger W_p=(1-p)I_C+pI_C=I_C$.
The pure output of the two channel dilations is
\[
 \ket{\Gamma}_{R_0 B K B' E'}
 =(I_{R_0}\otimes V_\sigma\otimes W_p)\ket{\Psi}_{R_0AC}.
\]
Bob receives $BB'$, and the complementary output is $KE'$.

To compute the two output entropies, denote the state immediately
before the action of $W_p$ by
\[
 \ket{\Xi_t}_{BKC}=(V_\sigma\otimes I_C)\ket{\psi_t}_{AC},
 \qquad
 \Theta_{BKC}=\tfrac12\proj{\Xi_0}+\tfrac12\proj{\Xi_1}.
\]
Tracing out $E'$ or $B'$ removes the cross terms between the two
branches of $W_p$. We order both direct sums below by the same
two events: with probability $1-p$, Bob receives $C$ and $E'$
contains the flag; with probability $p$, Bob receives the flag
and $E'$ receives $C$. Identifying the transmitted data space
with $C$ gives
\[
 \begin{aligned}
 \Gamma_{BB'}&\cong(1-p)\Theta_{BC}\oplus p\Theta_B,\\
 \Gamma_{KE'}&\cong(1-p)\Theta_K\oplus p\Theta_{KC}.
 \end{aligned}
\]
The direct-sum entropy formula
$S((1-p)\rho\oplus p\tau)=h_2(p)+(1-p)S(\rho)+pS(\tau)$
therefore yields
\[
 \begin{aligned}
 S(BB')_\Gamma&=h_2(p)+(1-p)S(BC)_\Theta+pS(B)_\Theta,\\
 S(KE')_\Gamma&=h_2(p)+(1-p)S(K)_\Theta+pS(KC)_\Theta.
 \end{aligned}
\]

For each $t$, tracing out $C$ in $\proj{\Xi_t}$ gives
$V_\sigma\rho_tV_\sigma^\dagger$. Its marginals on $B$ and $K$
are therefore the same $B_t$ and $K_t$ as in the private ensemble.
Since $\ket{\Xi_t}_{BKC}$ is pure, its $BC$ marginal has entropy
$S(K_t)$. The two $BC$ marginals are supported on the orthogonal
subspaces $B\otimes C_0$ and $B\otimes C_1$, respectively,
whereas $\Theta_K=(K_0+K_1)/2$. Hence
\[
 \begin{aligned}
 S(BC)_\Theta-S(K)_\Theta
 &=1+\tfrac12\bigl[S(K_0)+S(K_1)\bigr]
      -S\!\left(\tfrac{K_0+K_1}{2}\right)\\
 &=H(T|K)_\Omega ,
 \end{aligned}
\]
where the last equality uses Eq.~\eqref{sm:eq:model-holevo} for
the equiprobable ensemble. Interchanging $B$ and $K$ gives
$S(KC)_\Theta-S(B)_\Theta=H(T|B)_\Omega$.
Subtracting the two output entropies now gives, for $0\le p\le1$,
\begin{equation}
 \begin{aligned}
 I_c(\omega_{AC},\N_\sigma\otimes\E_p^{(32)})
 &=S(BB')_\Gamma-S(KE')_\Gamma\\
 &=(1-p)\bigl[S(BC)_\Theta-S(K)_\Theta\bigr]
    +p\bigl[S(B)_\Theta-S(KC)_\Theta\bigr]\\
 &=(1-p)H(T|K)_\Omega-pH(T|B)_\Omega .
 \end{aligned}
 \label{sm:eq:sm-erasure-general}
\end{equation}
Substituting the conditional entropies from
Eq.~\eqref{sm:eq:simple-conditional-entropies}, we obtain the
coherent information of this fixed input:
\begin{equation}
 f(p):=I_c(\omega_{AC},\N_\sigma\otimes\E_p^{(32)})
 =\frac{1-p}{40}-\frac{p\,h_2(1/5)}{32}.
 \label{sm:eq:helper-fp}
\end{equation}

For $p\ge1/2$, the erasure channel is antidegradable and has zero
quantum capacity~\cite{Bennett1997}. At $p=1/2$, we recover the
Smith--Yard relation~\cite{SmithYard2008}: the coherent information
is half the private information of the chosen ensemble. Thus,
although both individual channels have zero quantum capacity,
\begin{equation}
 Q(\N_\sigma\otimes\E_{1/2}^{(32)})
 \ge f(1/2)=\frac{\epsrate}{2}>0.
 \label{sm:eq:simple-erasure}
\end{equation}
The same input has positive coherent information at erasure
probabilities strictly above $1/2$. Indeed,
Eq.~\eqref{sm:eq:helper-fp} gives
\begin{equation}
 f(p)>0
 \quad\Longleftrightarrow\quad
 p<p_*:=\frac{4}{4+5h_2(1/5)}=0.525649\ldots .
 \label{sm:eq:helper-threshold}
\end{equation}
Thus $\omega_{AC}$ witnesses superactivation throughout
$\tfrac12\le p<p_*$. To obtain activation with a convolution
auxiliary, it therefore suffices to realize $\E_p^{(32)}$ for
one $p$ in this interval.

\subsection{The convolution auxiliary}
\label{sm:sec:convolution-helper}

The second auxiliary acts on a single seven-level system.
For $x,y\in\F_7$, take
\begin{equation}
 \begin{gathered}
 U_{\rm aux}\ket{x,y}=\ket{x+y,x-y},\qquad
 G_{\rm aux}=2^{-1}\begin{pmatrix}1&1\\1&-1\end{pmatrix},\\
 \ket{\xi}=\frac{\ket{1}+\ket{-1}}{\sqrt2},\qquad
 \mathcal A_7(\rho)
 =\Tr_{E_h}\!\left[
    U_{\rm aux}(\rho\otimes\proj{\xi})U_{\rm aux}^\dagger
   \right].
 \end{gathered}
 \label{sm:eq:helper-definition}
\end{equation}
Write $B_h$ and $E_h$ for the receiver and environment outputs.
Since the auxiliary environment $\ket{\xi}$ is pure, its Stinespring isometry is
\[
 W_7\ket{x}
 =\frac{\ket{x+1}_{B_h}\ket{x-1}_{E_h}
       +\ket{x-1}_{B_h}\ket{x+1}_{E_h}}{\sqrt2}.
\]
The right-hand side is invariant under exchange of $B_h$ and $E_h$
for every $x\in\F_7$. Thus, after identifying the two output spaces,
$\mathcal A_7=\mathcal A_7^c$. The same equality holds for every
tensor power, so $P(\mathcal A_7)=Q(\mathcal A_7)=0$.

To identify the qubit transmitted on a subspace, let
$\iota_7,j_7:\mathbb C^2\to\mathbb C^7$ be the input and
surviving-output embeddings, respectively:
\[
 \iota_7\ket{0}=\ket{1},\quad
 \iota_7\ket{1}=\ket{-1},\qquad
 j_7\ket{0}=\ket{2},\quad
 j_7\ket{1}=\ket{-2}.
\]
Evaluating $W_7$ at $x=1,-1$ gives, for any qubit vector $\ket{\psi}$,
\[
 W_7\iota_7\ket{\psi}
 =\frac{(j_7\ket{\psi})_{B_h}\ket{0}_{E_h}
        +\ket{0}_{B_h}(j_7\ket{\psi})_{E_h}}{\sqrt2}.
\]
Since $\ket{0}$ is orthogonal to
$j_7(\mathbb C^2)=\operatorname{span}\{\ket{2},\ket{-2}\}$,
tracing out $E_h$ yields, for every operator
$\theta$ on $\mathbb C^2$,
\begin{equation}
 \mathcal A_7(\iota_7\theta\iota_7^\dagger)
 =\tfrac12 j_7\theta j_7^\dagger
  +\tfrac12\Tr(\theta)\proj{0}.
 \label{sm:eq:helper-qubit}
\end{equation}
This is a qubit half-erasure channel: the surviving logical
$\ket{0}$ is the physical state $\ket{2}$, whereas the physical
state $\ket{0}$ is the erasure flag. The environment $\ket{\xi}$
is magic: its computational-basis support has size two, while a
pure stabilizer state of a single seven-level system has support
of size one or seven~\cite{Gross2006}.

The register $C$ in Eq.~\eqref{sm:eq:smomega} has dimension $32$.
Proposition~\ref{sm:prop:auxiliary-simulation} gives an encoder
and decoder that realize $\E_p^{(32)}$ on $C$ using $L$ copies of
$\mathcal A_7$, for some $\tfrac12<p<p_*$.
The encoded and decoded auxiliary acts exactly as $\E_p^{(32)}$
even when $C$ is entangled with a reference. Hence, for the input
$\omega_{AC}$, sending $A$ through $\N_\sigma$ and applying the
auxiliary code to $C$ gives the same coherent information $f(p)>0$
as the ideal joint channel in Eq.~\eqref{sm:eq:helper-fp}.
Section~\ref{sm:sec:theorem-two}
constructs the code and derives the resulting rate bound
$Q(\N_\sigma\otimes\mathcal A_7)\ge f(p)/L>0$.

\section{Private PPT convolution channels for prescribed couplings}
\label{sm:sec:theorem-one}\label{sm:sec:crossratio}\label{sm:sec:stabilizer-eb}

\begin{theorem}[Theorem 1 of the Letter and its consequences]
\label{sm:thm:crossratio}
Let $d\ge7$ be prime, let $n=7$, and put $D=d^7$.
Let $G\in\mathrm{GL}(2,\F_d)$ have all four entries nonzero.
Write
\[
 G^{-\mathsf T}=\begin{pmatrix}a&b\\c&e\end{pmatrix},
 \qquad r=\frac{ae}{bc},
\]
and suppose that $r\notin\{-1,1/2,1/3\}$.
There exists a mixed magic environment state $\sigma$ on
$(\mathbb C^d)^{\otimes7}$ such that
\begin{equation}
 Q(\N_{G,\sigma})=0,\qquad
 P(\N_{G,\sigma})\ge\epsrate,\qquad
 \epsrate=\frac{12/5-\log_2 5}{32}>0.
 \label{sm:eq:crossratio-capacities}
\end{equation}
The normalized Choi state $J_{G,\sigma}$ satisfies
\[
 J_{G,\sigma}^{T_B}\succeq0,\qquad
 \rank J_{G,\sigma}=288D,\qquad
 \rank J_{G,\sigma}^{T_B}=528D.
\]
Moreover,
\begin{equation}
 Q\bigl(\N_{G,\sigma}\otimes\E_{1/2}^{(32)}\bigr)
 \ge\epsrate/2>0.
 \label{sm:eq:crossratio-erasure}
\end{equation}
With $G$ and the half-erasure auxiliary held fixed, replacing
$\sigma$ by any $\tau\in\STAB_7$ makes the joint quantum capacity
zero.
\end{theorem}

\begin{proof}
\emph{Reduction and construction.}
For $\gamma\in\F_d^\times$, let $S_\gamma\ket{x}=\ket{\gamma x}$
for $x\in V=\F_d^7$. Set $h=-1/r$ and $\alpha=h+1=1-1/r$.
Since all four entries are nonzero and $ae-bc\ne0$, we have
$r\ne0,1$ and hence $\alpha\ne0,1$. The excluded values
$r=-1,1/2,1/3$ correspond to $\alpha=2,-1,-2$, respectively.

Let $U_h\ket{x,y}=\ket{x+y,y-hx}$, and denote by $\N_{h,\sigma}$
the channel it induces with environment $\sigma$. Direct
substitution gives
\begin{equation}
 \begin{aligned}
 U_h
 &=(S_{1/b}\otimes S_{1/e})U_G(S_{b/a}\otimes I_Y),\\
 \N_{h,\sigma}(\rho)
 &=S_{1/b}\,
   \N_{G,\sigma}\bigl(S_{b/a}\rho S_{b/a}^\dagger\bigr)\,
   S_{1/b}^\dagger .
 \end{aligned}
 \label{sm:eq:crossratio-normalization}
\end{equation}
The rescaling acts trivially on $Y$, so the environment state
is unchanged. For corresponding inputs $\rho$ and
$S_{b/a}\rho S_{b/a}^\dagger$, the two Bob outputs differ by
$S_{1/b}$. With the same purification of $\sigma$, the two
complementary outputs differ by $S_{1/e}\otimes I_F$.
The normalized Choi states satisfy
\[
 J_{h,\sigma}=LJ_{G,\sigma}L^\dagger,\qquad
 L=S_{b/a}^{\mathsf T}\otimes S_{1/b}.
\]
Since the rescalings are real permutation matrices,
$J_{h,\sigma}^{T_B}=LJ_{G,\sigma}^{T_B}L^\dagger$ as well.
It therefore suffices to compute the two Choi ranks and verify
the PPT property for $U_h$.

Choose a nonsquare $\nu\in\F_d$ and define
\[
 a_{4u+v}=(u,v,u^2-\nu v^2),\qquad u,v\in\{0,1,2,3\}.
\]
The subscript is an integer in $\{0,\ldots,15\}$, and the
coordinates are evaluated in $\F_d$. For fixed $d$, these
sixteen points can be used for every allowed coupling.
With $0_3=(0,0,0)$, set
\begin{equation}
 \begin{aligned}
 r_i&=(1,a_i,0_3),&s_j&=(0,0_3,a_j),\\
 u_{ij}&=-r_i-\alpha s_j,&v_{ij}&=-\alpha r_i-s_j,\\
 w_i&=-(\alpha+1)r_i,&z_j&=-(\alpha+1)s_j,
 \end{aligned}
 \qquad 0\le i,j\le15 .
 \label{sm:eq:crossratio-labels}
\end{equation}
Use the real Hadamard matrix $H_{16}$ and the constants $m=16$,
$Z=640$ from Sec.~\ref{sm:sec:example}. Define the unnormalized
vectors $\ket{\phi_{ij}}=\ket{u_{ij}}+(H_{16})_{ij}\ket{v_{ij}}$.
The environment and the two equiprobable signals are
\begin{equation}
 \begin{gathered}
 \sigma=\frac1Z\left[
   \sum_{i,j}\proj{\phi_{ij}}
   +4\sum_i\proj{w_i}+4\sum_j\proj{z_j}\right],\\
 \rho_0=\frac1m\sum_i\proj{r_i},\qquad
 \rho_1=\frac1m\sum_j\proj{s_j},\qquad
 \Pr(T=0)=\Pr(T=1)=\tfrac12 .
 \end{gathered}
 \label{sm:eq:crossratio-state}
\end{equation}
The coefficient $\alpha+1$ in the labels $w_i,z_j$ is a field
element determined by the coupling; the weight $4$ of their
diagonal terms in $\sigma$ is a fixed real number.
The first coordinates of $u_{ij},v_{ij},w_i,z_j$ are
$-1,-\alpha,-(\alpha+1),0$, respectively. They are distinct
since $\alpha\notin\{0,1,-1\}$. Within each family, the remaining
coordinates distinguish the indices. Thus all $2m^2+2m=544$
environment labels are distinct. Each pair contributes an
eigenvalue $2$ to $Z\sigma$, and each singleton contributes
an eigenvalue $4$. Hence $\Tr\sigma=(2m^2+8m)/Z=1$ and
$\rank\sigma=m^2+2m=288$. For $d=7$, $\nu=3$, and $\alpha=3$,
this gives the construction of Sec.~\ref{sm:sec:example}.

\emph{PPT and zero quantum capacity.}
Let $J_{h,\eta}$ denote the normalized Choi state of $\N_{h,\eta}$
for an arbitrary environment state $\eta$. Tracing out $E$
imposes $y-hx=y'-hx'=:e_0$, giving
\begin{equation}
 J_{h,\eta}
 =\frac1D\sum_{e_0,x,x'\in V}
   \eta_{e_0+hx,e_0+hx'}\,
   \ket{x,e_0+\alpha x}\bra{x',e_0+\alpha x'} .
 \label{sm:eq:canonical-choi-expansion}
\end{equation}
For fixed $e_0$, the relabeling $x\mapsto e_0+hx$ identifies the
corresponding block with $\eta/D$. After partial transpose on
$B$, the ket and bra instead share the label
$s=e_0+\alpha(x+x')$. In the basis
$\{\ket{x,s-\alpha x}:x\in V\}$, the $s$ block has entries
\[
 \frac1D\,\eta_{s-x-\alpha x',\,s-\alpha x-x'}.
\]
Relabel its rows and columns by $u=s-(\alpha+1)x$ and
$v=s-(\alpha+1)x'$. Substitution gives
\begin{equation}
 \begin{gathered}
 J_{h,\eta}\simeq\bigoplus_{e_0\in V}\frac{\eta}{D},
 \qquad
 J_{h,\eta}^{T_B}\simeq
 \bigoplus_{s\in V}\frac{\mathcal R_\alpha(\eta)}{D},\\
 [\mathcal R_\alpha(X)]_{u,v}
 =X_{(u+\alpha v)/(\alpha+1),\,(\alpha u+v)/(\alpha+1)} .
 \end{gathered}
 \label{sm:eq:canonical-choi-blocks}
\end{equation}
Both relabelings are bijective, since $h=\alpha-1\ne0$ and
$\alpha+1\ne0$. As in Sec.~\ref{sm:sec:example}, each
$\simeq$ denotes a simultaneous permutation of rows and columns.

The entry rearrangement $\mathcal R_\alpha$ fixes the diagonal
matrix units. On an arbitrary matrix unit it acts as
\[
 \mathcal R_\alpha(\ket{y}\bra{z})
 =\ket{(y-\alpha z)/(1-\alpha)}
  \bra{(z-\alpha y)/(1-\alpha)} .
\]
For the labels in Eq.~\eqref{sm:eq:crossratio-labels},
\[
 \frac{u_{ij}-\alpha v_{ij}}{1-\alpha}=w_i,\qquad
 \frac{v_{ij}-\alpha u_{ij}}{1-\alpha}=z_j.
\]
Thus $\ket{u_{ij}}\bra{v_{ij}}$ moves to $\ket{w_i}\bra{z_j}$,
and its adjoint moves to $\ket{z_j}\bra{w_i}$. Consequently,
\begin{equation}
 Z\mathcal R_\alpha(\sigma)\simeq
 I_{512}\oplus
 \begin{pmatrix}
  4I_{16}&H_{16}\\ H_{16}^{\mathsf T}&4I_{16}
 \end{pmatrix}
 \oplus0 .
 \label{sm:eq:crossratio-block}
\end{equation}
The $512$- and $32$-dimensional blocks coincide with those in
Eq.~\eqref{sm:eq:simple-pt-block}; the remaining block is zero.
Hence $\mathcal R_\alpha(\sigma)$ is positive semidefinite and
has rank $512+16=528$. Equation~\eqref{sm:eq:canonical-choi-blocks}
now gives
\[
 J_{h,\sigma}^{T_B}\succeq0,\qquad
 \rank J_{h,\sigma}=288D,\qquad
 \rank J_{h,\sigma}^{T_B}=528D .
\]
These conclusions carry over to $J_{G,\sigma}$ by the local
rescalings above. The PPT property therefore gives
$Q(\N_{G,\sigma})=0$~\cite{Horodecki1998,Horodecki2000Binding}.

\emph{Positive private information.}
Use the purification in Eq.~\eqref{sm:eq:simple-purification}
with the labels from Eq.~\eqref{sm:eq:crossratio-labels}:
$u_{ij}$ and $v_{ij}$ share the pair flag $ij$, and each $w_i$
and $z_j$ has its own flag. Let $\mathcal Y$ denote these $544$
environment basis labels. For a computational-basis input
$\ket{x}$, the two marginal states are
\begin{equation}
 \begin{aligned}
 B(x)&=\frac1Z\sum_{y\in\mathcal Y}q_y\proj{x+y},\\
 K(x)&=\frac1Z\sum_{y\in\mathcal Y}q_y\proj{y-hx,f(y)},
 \qquad K=EF,
 \end{aligned}
 \label{sm:eq:crossratio-marginals}
\end{equation}
where $q_{u_{ij}}=q_{v_{ij}}=1$ and $q_{w_i}=q_{z_j}=4$.
Indeed, for fixed $x$, both $y\mapsto x+y$ and $y\mapsto y-hx$
are injective, so the partial traces remove the cross terms.
For the ensemble in Eq.~\eqref{sm:eq:crossratio-state}, each
input component and environment label contribute a joint weight
$q_y/(2mZ)$. By Eq.~\eqref{sm:eq:conditional-entropy-overlap},
only outputs shared by the two signals contribute to the
conditional entropies.

For Bob, the first coordinate of $x+y$ is given by
\[
 \begin{array}{c|rrrr}
     &u_{ij}&v_{ij}&w_i&z_j\\ \hline
 T=0 &0&1-\alpha&-\alpha&1\\
 T=1 &-1&-\alpha&-(\alpha+1)&0
 \end{array}
\]
The two rows share $0$ and $-\alpha$. Since
$\alpha\notin\{0,1,-1\}$, the only possible additional matches
are $1-\alpha=-1$ and $1=-(\alpha+1)$, which would require
$\alpha=2$ and $\alpha=-2$, respectively. Both are excluded.

To compare the remaining coordinates, we use
\begin{equation}
 a_i+\alpha a_j=(\alpha+1)a_\ell
 \quad\Longrightarrow\quad i=j=\ell .
 \label{sm:eq:crossratio-code}
\end{equation}
Write $a_i=(\xi,q(\xi))$, $a_j=(\eta,q(\eta))$, and
$a_\ell=(\zeta,q(\zeta))$, where
$q:\F_d^2\to\F_d$ is given by $q(\xi)=\xi_1^2-\nu\xi_2^2$.
The first two coordinates give
$\zeta=(\xi+\alpha\eta)/(\alpha+1)$, and the last gives
\[
 \begin{aligned}
 0
 &=q(\xi)+\alpha q(\eta)
   -(\alpha+1)q\!\left(\frac{\xi+\alpha\eta}{\alpha+1}\right)\\
 &=\frac{\alpha}{\alpha+1}q(\xi-\eta).
 \end{aligned}
\]
The form $q$ vanishes only at zero: if $q(\delta)=0$ and
$\delta_2\ne0$, then $\nu=(\delta_1/\delta_2)^2$, contrary
to the choice of $\nu$; if $\delta_2=0$, then $\delta_1=0$.
Thus $\xi=\eta=\zeta$, proving Eq.~\eqref{sm:eq:crossratio-code}.

At first coordinate $0$ and $-\alpha$, respectively, equality
of Bob's outputs reduces to
\[
 \begin{aligned}
 r_\kappa+u_{ij}=s_\ell+z_{j'}
 &\quad\Longleftrightarrow\quad
 a_\kappa=a_i,\quad a_\ell+\alpha a_j=(\alpha+1)a_{j'},\\
 r_\kappa+w_i=s_\ell+v_{i'j}
 &\quad\Longleftrightarrow\quad
 a_\ell=a_j,\quad a_\kappa+\alpha a_{i'}=(\alpha+1)a_i .
 \end{aligned}
\]
Equation~\eqref{sm:eq:crossratio-code} forces
$\kappa=i$ and $\ell=j=j'$ in the first line, and
$\ell=j$ and $\kappa=i'=i$ in the second. Bob's complete list
of ambiguous outputs, with weights in units of $1/(2mZ)$, is
therefore
\[
 \begin{array}{c|cc}
 \beta&2mZ\Pr(T=0,B=\beta)&2mZ\Pr(T=1,B=\beta)\\ \hline
 -\alpha s_j&m&4\\
 -\alpha r_i&4&m
 \end{array}
\]
Each row represents $m$ distinct outputs. For fixed $j$, all
$m$ terms $r_i+u_{ij}$ give $-\alpha s_j$; for fixed $i$, all
$m$ terms $s_j+v_{ij}$ give $-\alpha r_i$.

For Eve, equality of outputs requires equality of both the
$E$ coordinate and the flag in $F$. With a singleton flag, or
with the same endpoint of a pair flag, a common output would
require $y-hr_\kappa=y-hs_\ell$. Since $h\ne0$, this would give
$r_\kappa=s_\ell$, contrary to their first coordinates.

For opposite endpoints of the same flag $ij$, note that
\[
 u_{ij}-v_{ij}=(\alpha-1)(r_i-s_j)=h(r_i-s_j).
\]
It follows that
\[
 u_{ij}-hr_\kappa=v_{ij}-hs_\ell
 \quad\Longleftrightarrow\quad
 r_i-r_\kappa=s_j-s_\ell .
\]
The two differences are supported on disjoint coordinate blocks,
so they both vanish. Hence $\kappa=i$ and $\ell=j$, giving the
common output $(-\alpha(r_i+s_j),ij)$. The reverse orientation
would instead require $r_i+r_\kappa=s_j+s_\ell$, whose first
coordinates would give $2=0$ in $\F_d$. Thus Eve has exactly
$m^2$ ambiguous outputs, each with weights $(1,1)$ in units
of $1/(2mZ)$.

Substitution in Eq.~\eqref{sm:eq:conditional-entropy-overlap}
gives
\begin{equation}
 \begin{aligned}
 H(T|B)
 &=2m\,\frac{m+4}{2mZ}\,
   \hb\!\left(\frac{4}{m+4}\right)
   =\frac{\hb(1/5)}{32},\\
 H(T|K)
 &=m^2\,\frac{2}{2mZ}\,\hb(1/2)
   =\frac1{40}.
 \end{aligned}
 \label{sm:eq:crossratio-entropies}
\end{equation}
The private information of this ensemble is therefore
$\Delta=H(T|K)-H(T|B)=\epsrate>0$.

\emph{Superactivation and magic.}
For the canonical interaction $U_h$, purify $\rho_0,\rho_1$
in orthogonal $16$-dimensional sectors of $C$ as in
Eq.~\eqref{sm:eq:smpsit}, using the present $r_i,s_j$.
Let $\omega_{AC}$ be the equiprobable mixture of the two
purifications. Equations~\eqref{sm:eq:sm-erasure-general}
and~\eqref{sm:eq:crossratio-entropies} give
\begin{equation}
 I_c\bigl(\omega_{AC},\N_{h,\sigma}\otimes\E_p^{(32)}\bigr)
 =\frac{1-p}{40}-\frac{p\,\hb(1/5)}{32}
 =f(p).
 \label{sm:eq:crossratio-coherent-information}
\end{equation}

For the prescribed coupling $G$, use the signal states
$\rho_t^G=S_{b/a}\rho_tS_{b/a}^\dagger$ and the joint input
\[
 \omega^G_{AC}
 =(S_{b/a}\otimes I_C)\,\omega_{AC}\,
  (S_{b/a}^\dagger\otimes I_C).
\]
By Eq.~\eqref{sm:eq:crossratio-normalization}, the corresponding
Bob and complementary outputs differ from those for $U_h$ only
by local unitaries. The ensemble $\{\tfrac12,\rho_t^G\}_{t=0,1}$
therefore has private information $\epsrate$, giving
$P(\N_{G,\sigma})\ge\epsrate$. Likewise,
\[
 I_c\bigl(\omega^G_{AC},\N_{G,\sigma}\otimes\E_p^{(32)}\bigr)
 =f(p).
\]
At $p=1/2$, this gives $f(1/2)=\epsrate/2>0$ and proves
Eq.~\eqref{sm:eq:crossratio-erasure}. The same input witnesses
superactivation throughout $1/2\le p<p_*$, with $p_*$ as in
Eq.~\eqref{sm:eq:helper-threshold}.

For a positive coupling on $k$ qudits, the proof of Theorem~1
in Ref.~\cite{Xiong2026} shows that a pure stabilizer environment
induces an entanglement-breaking channel. The induced channel
depends affinely on the environment state by
Eq.~\eqref{sm:eq:model-convolution}. Since entanglement-breaking
channels form a convex set, it follows that
\begin{equation}
 \tau\in\STAB_k
 \quad\Longrightarrow\quad
 \N_{G,\tau}\ \text{is entanglement breaking}.
 \label{sm:eq:stabilizer-eb}
\end{equation}
Such a channel has zero private capacity. Since
$P(\N_{G,\sigma})\ge\epsrate>0$, the environment constructed
above lies outside $\STAB_7$ and is therefore magic.
For any replacement $\tau\in\STAB_7$, the channel $\N_{G,\tau}$
is entanglement breaking and hence antidegradable. The half-erasure
channel is also antidegradable, so their product is antidegradable.
Consequently,
\begin{equation}
 Q\bigl(\N_{G,\tau}\otimes\E_{1/2}^{(32)}\bigr)=0
 \qquad(\tau\in\STAB_7).
 \label{sm:eq:smfreecomparison}
\end{equation}
\end{proof}

\section{Superactivation with a convolution auxiliary}
\label{sm:sec:two-convolution}\label{sm:sec:theorem-two}

\subsection{The physical auxiliary and its qubit restriction}

We use the auxiliary of Sec.~\ref{sm:sec:convolution-helper}
in any prime dimension $d\ge7$. For $x,y\in\F_d$, define
\begin{equation}
 \begin{gathered}
 U_{\rm aux}\ket{x,y}=\ket{x+y,x-y},\qquad
 G_{\rm aux}=2^{-1}\begin{pmatrix}1&1\\1&-1\end{pmatrix},\\
 \ket{\xi}=\frac{\ket{1}+\ket{-1}}{\sqrt2},\qquad
 \mathcal A_d(\rho)=\Tr_{E_h}\!\left[
   U_{\rm aux}(\rho\otimes\proj{\xi})U_{\rm aux}^\dagger
 \right].
 \end{gathered}
 \label{sm:eq:theorem-two-helper}
\end{equation}
Each use couples one input qudit to a fresh copy of $\ket{\xi}$.
Bob receives $B_h$. Since the environment input is pure, $E_h$
is the full complementary output. The Stinespring isometry is
\begin{equation}
 W_d\ket{x}
 =U_{\rm aux}(\ket{x}\otimes\ket{\xi})
 =\frac{\ket{x+1}_{B_h}\ket{x-1}_{E_h}
       +\ket{x-1}_{B_h}\ket{x+1}_{E_h}}{\sqrt2}.
 \label{sm:eq:auxiliary-isometry}
\end{equation}
The right-hand side is invariant under exchange of $B_h$ and
$E_h$ for every $x$. After identifying the two output spaces,
$\mathcal A_d=\mathcal A_d^c$. The two outputs thus agree for
every input, also at each tensor power. Both the private and
coherent informations vanish, so $P(\mathcal A_d)=Q(\mathcal A_d)=0$.

Let $\iota_d,j_d:\mathbb C^2\to\mathbb C^d$ be the input and
surviving-output embeddings, respectively:
\[
 \iota_d\ket{0}=\ket{1},\quad
 \iota_d\ket{1}=\ket{-1},\qquad
 j_d\ket{0}=\ket{2},\quad
 j_d\ket{1}=\ket{-2}.
\]
Evaluating Eq.~\eqref{sm:eq:auxiliary-isometry} at $x=1,-1$
and extending linearly gives the first identity below for any
qubit vector $\ket{\psi}$. Tracing out $E_h$ gives the second
for every operator $X\in\mathcal L(\mathbb C^2)$:
\begin{equation}
 \begin{aligned}
 W_d\iota_d\ket{\psi}
 &=\frac{(j_d\ket{\psi})_{B_h}\ket{0}_{E_h}
         +\ket{0}_{B_h}(j_d\ket{\psi})_{E_h}}{\sqrt2},\\
 \mathcal A_d(\iota_d X\iota_d^\dagger)
 &=\tfrac12 j_d Xj_d^\dagger
   +\tfrac12\Tr(X)\proj{0}_{B_h}.
 \end{aligned}
 \label{sm:eq:auxiliary-erasure-restriction}
\end{equation}
The cross terms vanish in the partial trace because the physical
flag $\ket{0}$ is orthogonal to
$j_d(\mathbb C^2)=\operatorname{span}\{\ket{2},\ket{-2}\}$.
Thus the restriction is a qubit half-erasure channel. In
particular, a transmitted logical $\ket{0}$ arrives as $\ket{2}$,
whereas an erasure gives $\ket{0}$.

The code below applies $\iota_d$ to each physical code qubit
before one use of $\mathcal A_d$. Since
Eq.~\eqref{sm:eq:auxiliary-erasure-restriction} holds for every
qubit operator, it also applies when these qubits are entangled
with one another or with a reference.

\subsection{An explicit simulation of a 32-dimensional erasure channel}

\begin{proposition}[Simulation by a finite convolution code]
\label{sm:prop:auxiliary-simulation}
For each prime $d\ge7$, the channel $\mathcal A_d$ admits an
isometric encoder $V_d:C\to(\mathbb C^d)^{\otimes L}$,
$\dim C=32$, and a completely positive trace-preserving decoder
$\mathcal D_d$ with output space
$\widehat C\oplus\mathbb C\ket{\mathrm{er}}$, where
$\widehat C\simeq C$, such that
\begin{equation}
 \begin{gathered}
 \mathcal D_d\circ\mathcal A_d^{\otimes L}\circ\mathcal V_d
 =\mathcal E_p^{(32)},\qquad
 \mathcal V_d(X)=V_dXV_d^\dagger,\\
 L=1\,769\,472,\qquad
 \frac12<p<p_0:=\frac{33823}{65536}.
 \end{gathered}
 \label{sm:eq:theorem-two-logical}
\end{equation}
The encoder and decoder do not depend on the channel paired
with $\mathcal A_d$.
\end{proposition}

\begin{proof}
\emph{Encoding the logical register.}
Identify $C$ with five logical qubits and identify $\widehat C$
with $C$ in the computational basis. Take integers $m_c\ge4$
and $r_c\ge1$, and set
\[
 N=2m_c,\qquad t_c=m_c-3.
\]
The data block has $N$ qubits. A separate header records which
encoding was used, with each of its $N$ label bits repeated
$r_c$ times. We choose $m_c$ and $r_c$ after deriving the
success probability. All coding matrices and bit strings below
are over $\mathbb F_2$.

Fix an ordered $\mathbb F_2$-basis $(\beta_1,\ldots,\beta_N)$
of $\mathbb K=\mathbb F_{2^N}$, and write $[x]$ for the
coordinate column of $x\in\mathbb K$. For each nonzero
$\lambda\in\mathbb K$, multiplication by $\lambda$ has matrix
\[
 \mathsf M_\lambda
 =\bigl([\lambda\beta_1]\ \cdots\ [\lambda\beta_N]\bigr),
 \qquad
 [\lambda x]=\mathsf M_\lambda[x].
\]
Its inverse is $\mathsf M_{\lambda^{-1}}$. Let $\mathsf A_\lambda$
consist of columns $1,\ldots,t_c$ of $\mathsf M_\lambda$, and
let $\mathsf B_\lambda$ consist of columns $t_c+1,\ldots,t_c+5$
of the same matrix. Thus $\mathsf A_\lambda$ has size
$N\times t_c$ and $\mathsf B_\lambda$ has size $N\times5$.

For fixed $\lambda$, prepare $t_c$ ancillary qubits in
$\ket{+}^{\otimes t_c}$ and $N-t_c-5$ in
$\ket{0}^{\otimes(N-t_c-5)}$. Apply the basis permutation
$P_\lambda\ket{\mathbf x}=\ket{\mathsf M_\lambda\mathbf x}$
to the ordered registers $U_{\rm in},C,Z_{\rm in}$. For a
logical basis input $\ell\in\mathbb F_2^5$, this gives
\begin{equation}
 \begin{aligned}
 \ket{+}^{\otimes t_c}_{U_{\rm in}}
 \ket{\ell}_C
 \ket{0}^{\otimes(N-t_c-5)}_{Z_{\rm in}}
 &\xrightarrow{\ P_\lambda\ } V_\lambda\ket{\ell},\\
 V_\lambda\ket{\ell}
 &=2^{-t_c/2}\sum_{u\in\mathbb F_2^{t_c}}
   \ket{\mathsf A_\lambda u+\mathsf B_\lambda\ell}_{D_{\rm data}} .
 \end{aligned}
 \label{sm:eq:helper-data-encoder}
\end{equation}
The three input registers together become the $N$-qubit data
block $D_{\rm data}$. The sum over $u$ is coherent and includes
$u=0$. Its basis labels are all distinct as $(u,\ell)$ varies:
each is the image under the invertible matrix $\mathsf M_\lambda$
of the column with successive blocks $u,\ell,0$. It follows that
$V_\lambda^\dagger V_\lambda=I_C$. For fixed $\lambda$,
$P_\lambda$ can be implemented with CNOT and SWAP gates by
binary row reduction.

Write $[\lambda]=(\lambda_1,\ldots,\lambda_N)^{\mathsf T}$ and
define the header string
\[
 h_\lambda=\operatorname{rep}_{r_c}([\lambda])
 =(\underbrace{\lambda_1,\ldots,\lambda_1}_{r_c},\ldots,
   \underbrace{\lambda_N,\ldots,\lambda_N}_{r_c}).
\]
The full binary encoder and the physical encoder are
\begin{equation}
 \begin{aligned}
 V_{\rm bin}\ket{\psi}
 &=\frac1{\sqrt{2^N-1}}\sum_{\lambda\in\mathbb K^\times}
   \ket{h_\lambda}_H\otimes V_\lambda\ket{\psi},\\
 L&=(r_c+1)N,\qquad V_d=\iota_d^{\otimes L}V_{\rm bin}.
 \end{aligned}
 \label{sm:eq:helper-full-encoder}
\end{equation}
Here $H$ contains the $r_cN$ header qubits; the five logical
qubits are encoded in the data block. Distinct $\lambda$ give
orthogonal header states, so
\[
 V_{\rm bin}^\dagger V_{\rm bin}
 =\frac1{2^N-1}\sum_{\lambda\in\mathbb K^\times}
   V_\lambda^\dagger V_\lambda=I_C.
\]

To implement Eq.~\eqref{sm:eq:helper-full-encoder}, prepare the
$N$-qubit label register in
\[
 \ket{\Lambda}
 =\frac1{\sqrt{2^N-1}}\sum_{\lambda\in\mathbb K^\times}
   \ket{[\lambda]} .
\]
For example, the reflection $I-2\proj{v}$ maps $\ket{0^N}$
to $\ket{\Lambda}$ for
$\ket{v}=(\ket{0^N}-\ket{\Lambda})/\sqrt2$.
Apply $P_\lambda$ to the data inputs controlled by the label
$[\lambda]$, taking the controlled operation to be the identity
on the unused zero-label sector. For each label bit, use CNOT
gates to write its computational-basis value into $r_c-1$ fresh
zero qubits. The original $N$ label qubits and the added qubits
form $H$ and are all transmitted.

\emph{Transmission and acceptance tests.}
All $L$ positions pass through independent uses of $\mathcal A_d$.
At each output Bob measures the three projectors
\[
 j_dj_d^\dagger,\qquad
 \proj{0}_{B_h},\qquad
 I_{B_h}-j_dj_d^\dagger-\proj{0}_{B_h}.
\]
The first outcome identifies a surviving qubit without resolving
its computational-basis value. The second identifies a physical
erasure. The third has zero probability on encoded inputs and
is rejected. Bob identifies each surviving data subspace with
a qubit by $j_d^\dagger$.

Let $\eta\subseteq[L]$ be the set of erased positions in the
entire transmitted block, where $[L]=\{1,\ldots,L\}$.
By Eq.~\eqref{sm:eq:auxiliary-erasure-restriction}, the effects
of survival and erasure on a binary input are both $I_2/2$.
Thus a specified pattern $\eta$ has effect $2^{-L}I_{2^L}$ on
the binary block and effect
\begin{equation}
 V_{\rm bin}^\dagger(2^{-L}I_{2^L})V_{\rm bin}=2^{-L}I_C
 \label{sm:eq:helper-pattern-probability}
\end{equation}
on the logical input. Every pattern has probability $2^{-L}$,
including for inputs entangled with a reference. After the fixed
physical flags are omitted and each surviving subspace is identified
with a qubit, the unnormalized branch map on the binary block is
$X\mapsto2^{-L}\Tr_\eta X$.

Bob accepts the header if every group of $r_c$ copies has at
least one survivor and the surviving copies identify a consistent
nonzero label. He reads these copies in the computational basis.
Inconsistent copies and the zero label do not occur on encoded
inputs. Let $\mathsf E\subseteq[N]$ be the erased positions
\emph{in the data block}, let $\mathsf S=[N]\setminus\mathsf E$
be the surviving positions, and put $e=|\mathsf E|$.
A subscript $\mathsf E$ or $\mathsf S$ on a coding matrix
selects the corresponding rows in increasing order.

Given the recovered label $\lambda$, Bob accepts the data if
\begin{equation}
 e\le m_c-4,\qquad
 \rank\mathsf A_{\lambda,\mathsf E}=e,\qquad
 \rank[\,\mathsf A_{\lambda,\mathsf S}\ \
          \mathsf B_{\lambda,\mathsf S}\,]=t_c+5.
 \label{sm:eq:helper-correctability}
\end{equation}
Here $m_c=N/2$ is the mean number of erased data positions.
Denote the family of accepted data erasure sets by
\[
 \mathcal G_\lambda
 :=\{\mathsf E\subseteq[N]:
       \mathsf E\text{ satisfies Eq.~\eqref{sm:eq:helper-correctability}}\}.
\]
Bob knows $\mathsf E$ from the physical flags and computes both
ranks by binary Gaussian elimination.

To determine the state on which the recovery acts, fix a readable
header pattern and a data erasure set $\mathsf E$. Write
$a_\lambda,b_\lambda$ for the surviving and erased parts of
$h_\lambda$, respectively. For a reference--input state
$\theta_{RC}$, put
\[
 X_{\lambda\mu}
 =(I_R\otimes V_\lambda)\theta_{RC}
  (I_R\otimes V_\mu^\dagger).
\]
Conditional on the specified erasure pattern, tracing out the
lost positions in Eq.~\eqref{sm:eq:helper-full-encoder} gives
the normalized state
\[
 \frac1{2^N-1}\sum_{\lambda,\mu\in\mathbb K^\times}
 \langle b_\mu\mid b_\lambda\rangle
 \ket{a_\lambda}\bra{a_\mu}\otimes
 \Tr_{D_{\mathsf E}}X_{\lambda\mu},
\]
with the fixed physical flags omitted. Since the header is
readable, the strings $a_\lambda$ are distinct. For the
measurement outcome identifying $\lambda_0$, only the term
$\lambda=\mu=\lambda_0$ remains. The resulting unnormalized
state on $R D_{\mathsf S}$ is
\begin{equation}
 \frac1{2^N-1}\Tr_{D_{\mathsf E}}\!\left[
 (I_R\otimes V_{\lambda_0})\theta_{RC}
 (I_R\otimes V_{\lambda_0}^\dagger)\right].
 \label{sm:eq:helper-damaged-header}
\end{equation}
Its trace is $1/(2^N-1)$. Thus, for each specified pattern with
a readable header, the label is uniform before the data tests
are applied. Conditioned on that label, Bob has the reduced
data state for the fixed encoder $V_{\lambda_0}$.

\emph{Recovery on the surviving data.}
Fix an accepted pair $(\lambda,\mathsf E)$ and suppress the
index $\lambda$ temporarily. Since
$[\,\mathsf A_{\mathsf S}\ \mathsf B_{\mathsf S}\,]$
has full column rank, complete it to an invertible matrix
\[
 \mathsf T
 =[\,\mathsf A_{\mathsf S}\ \mathsf B_{\mathsf S}\ \mathsf W\,]
 \in\mathrm{GL}(N-e,\mathbb F_2).
\]
For a definite choice of $\mathsf W$, append, in order, each
standard basis column that increases the rank, stopping at
rank $N-e$. On a surviving data label,
\[
 \mathsf T^{-1}(\mathsf A_{\mathsf S}u+\mathsf B_{\mathsf S}\ell)
 =\begin{pmatrix}u\\ \ell\\ 0\end{pmatrix}.
\]

Next, select the first $e$ linearly independent columns of
$\mathsf A_{\mathsf E}$, scanning from left to right, and denote
their index set by $\mathcal J$. Write $\mathsf A_{\mathsf E,\mathcal J}$
for the resulting $e\times e$ matrix. Let
$\mathsf P_{\mathcal J}:\mathbb F_2^e\to\mathbb F_2^{t_c}$
insert the entries of a column at the positions in $\mathcal J$
and zeros elsewhere. Define
\[
 \mathsf R
 =\mathsf P_{\mathcal J}\mathsf A_{\mathsf E,\mathcal J}^{-1},
 \qquad
 \mathsf D=\mathsf R\mathsf B_{\mathsf E}.
\]
Then $\mathsf A_{\mathsf E}\mathsf R=I_e$ and hence
$\mathsf A_{\mathsf E}\mathsf D=\mathsf B_{\mathsf E}$.
For $e=0$, take $\mathsf D=0$.

Restoring the indices, Bob applies the following two basis
permutations to the $N-e$ surviving data qubits:
\begin{equation}
 \begin{aligned}
 U_{\lambda,\mathsf E}\ket{y}
   &=\ket{\mathsf T_{\lambda,\mathsf E}^{-1}y},\\
 S_{\lambda,\mathsf E}\ket{u,\ell,z}
   &=\ket{u+\mathsf D_{\lambda,\mathsf E}\ell,\ell,z},\\
 \mathcal R_{\lambda,\mathsf E}(X)
   &=\Tr_{UZ}\!\left[
       S_{\lambda,\mathsf E}U_{\lambda,\mathsf E}X
       U_{\lambda,\mathsf E}^\dagger S_{\lambda,\mathsf E}^\dagger
      \right].
 \end{aligned}
 \label{sm:eq:helper-pattern-recovery}
\end{equation}
After the first permutation, the successive registers are
$U,\widehat C,Z$, of sizes $t_c$, $5$, and $N-e-t_c-5$,
respectively. The first permutation can be implemented by
CNOT and SWAP gates. For the second, each nonzero entry
$(\mathsf D_{\lambda,\mathsf E})_{ij}$ specifies a CNOT
from qubit $j$ of $\widehat C$ to qubit $i$ of $U$.

Let $\ket{\Psi}_{RC}=\sum_\ell\ket{\varphi_\ell}_R\ket{\ell}_C$
be an arbitrary pure reference--input state. For the accepted
pair $(\lambda,\mathsf E)$, a purification of the conditional
state on $R D_{\mathsf S}$ is obtained by retaining the erased
data in a register $E_{\rm loss}$:
\[
 2^{-t_c/2}\sum_{u,\ell}\ket{\varphi_\ell}_R
 \ket{\mathsf A_{\mathsf E}u+\mathsf B_{\mathsf E}\ell}_{E_{\rm loss}}
 \ket{\mathsf A_{\mathsf S}u+\mathsf B_{\mathsf S}\ell}_{D_{\mathsf S}} .
\]
The first permutation in Eq.~\eqref{sm:eq:helper-pattern-recovery}
changes the last ket to $\ket{u}_U\ket{\ell}_{\widehat C}\ket{0}_Z$.
The erased data still carry the label
$\mathsf A_{\mathsf E}u+\mathsf B_{\mathsf E}\ell$.
After the second permutation, put $w=u+\mathsf D\ell$. Since
addition is binary,
\[
 \mathsf A_{\mathsf E}u+\mathsf B_{\mathsf E}\ell
 =\mathsf A_{\mathsf E}(w+\mathsf D\ell)+\mathsf B_{\mathsf E}\ell
 =\mathsf A_{\mathsf E}w .
\]
For each $\ell$, the change from $u$ to $w$ permutes all of
$\mathbb F_2^{t_c}$. The joint state therefore factors as
\begin{equation}
 \ket{\Psi}_{R\widehat C}\otimes
 \left(2^{-t_c/2}\sum_{w\in\mathbb F_2^{t_c}}
       \ket{w}_U
       \ket{\mathsf A_{\lambda,\mathsf E}w}_{E_{\rm loss}}\right)
 \otimes\ket{0}_Z .
 \label{sm:eq:helper-recovery-factorization}
\end{equation}
The middle factor is normalized because the states $\ket{w}_U$
are orthogonal. Bob keeps $\widehat C$ and discards $U,Z$.
Neither permutation acts on $E_{\rm loss}$, so taking its partial
trace before the recovery gives the same reduced output on
$R\widehat C$. This proves exact recovery for every pure input
with a reference, and hence, by linearity, for every $\theta_{RC}$.

\emph{The complete decoded channel.}
On rejection, Bob discards his remaining systems and prepares
the logical erasure flag $\ket{\mathrm{er}}\perp\widehat C$.
On acceptance, he applies $\mathcal R_{\lambda,\mathsf E}$.
He discards the header and all classical records in either case.
These operations define a trace-preserving decoder on the entire
physical output space, including outcomes that have zero
probability on encoded inputs.

Each accepted pair consisting of a full erasure pattern and a
measured label has probability $2^{-L}/(2^N-1)$. Write $s$ for
their total probability. The accepted output on $R\widehat C$
is $s\theta_{R\widehat C}$, where $\widehat C$ is identified
with $C$. The rejected output has the form
$\tau_R^{\rm fail}\otimes\proj{\mathrm{er}}$, with
$\tau_R^{\rm fail}$ unnormalized. Since the entire operation
is local and trace preserving on the channel input side of the
cut $R:C$, the total reference marginal is $\theta_R$. Subtracting
the accepted marginal $s\theta_R$ gives
$\tau_R^{\rm fail}=(1-s)\theta_R$. Thus
\begin{equation}
 \bigl(\operatorname{id}_R\otimes
   (\mathcal D_d\circ\mathcal A_d^{\otimes L}\circ\mathcal V_d)\bigr)
   (\theta_{RC})
 =s\theta_{R\widehat C}\oplus
   (1-s)\theta_R\otimes\proj{\mathrm{er}} .
 \label{sm:eq:helper-reference-identity}
\end{equation}
It remains to bound the logical erasure probability $p=1-s$.

\emph{Success probability.}
A header bit is unreadable exactly when all $r_c$ copies are
erased. The $N$ groups occupy disjoint positions, so the header
success probability is
\begin{equation}
 q_H=(1-2^{-r_c})^N\ge1-N2^{-r_c}.
 \label{sm:eq:helper-header-success}
\end{equation}
For a fixed label $\lambda$, each data erasure set has probability
$2^{-N}$. Its data acceptance probability is therefore
\[
 s_\lambda=2^{-N}|\mathcal G_\lambda|.
\]
By Eq.~\eqref{sm:eq:helper-damaged-header}, the label read from
any readable header is uniform for every fixed data pattern.
The mean data acceptance probability and the total success
probability are thus
\[
 \bar s=\frac1{2^N-1}\sum_{\lambda\in\mathbb K^\times}s_\lambda,
 \qquad s=q_H\bar s .
\]

For a fixed nonzero binary column $v$, the vectors
$\mathsf M_\lambda v$ run once through all nonzero columns as
$\lambda$ ranges over $\mathbb K^\times$: their field elements
are the products of $\lambda$ with the nonzero element represented
by $v$. The same property holds for a nonzero row $w$. Indeed,
if $w\mathsf M_\lambda=w\mathsf M_\mu$ for distinct $\lambda,\mu$,
then $w\mathsf M_{\lambda+\mu}=0$. This is impossible because
$\lambda+\mu\ne0$ and $\mathsf M_{\lambda+\mu}$ is invertible.
The row map is therefore injective between two sets of $2^N-1$
elements and hence is bijective.

Fix $\mathsf E\subseteq[N]$ with $|\mathsf E|=e$.
In the next two estimates, only $\lambda$ is random, uniformly
in $\mathbb K^\times$. The surviving matrix fails to have full
column rank precisely when there is a nonzero kernel vector
with blocks $\delta u\in\mathbb F_2^{t_c}$ and
$\delta\ell\in\mathbb F_2^5$. In fact,
\[
 [\,\mathsf A_{\lambda,\mathsf S}\ \
    \mathsf B_{\lambda,\mathsf S}\,]
 \begin{pmatrix}\delta u\\ \delta\ell\end{pmatrix}=0
 \quad\Longleftrightarrow\quad
 \operatorname{supp}\!\left[
   \mathsf M_\lambda
   \begin{pmatrix}\delta u\\ \delta\ell\\ 0\end{pmatrix}
 \right]\subseteq\mathsf E .
\]
There are $2^{t_c+5}-1$ nonzero candidate columns. For each one,
its image is uniform over the $2^N-1$ nonzero columns, of which
$2^e-1$ have support contained in $\mathsf E$. A union bound gives
\begin{equation}
 \begin{aligned}
 \Pr_\lambda\!\left\{
   \rank[\,\mathsf A_{\lambda,\mathsf S}\ \
            \mathsf B_{\lambda,\mathsf S}\,]<t_c+5
 \right\}
 &\le\frac{(2^{t_c+5}-1)(2^e-1)}{2^N-1}\\
 &<2^{t_c+5+e-N}.
 \end{aligned}
 \label{sm:eq:helper-surviving-rank-bound}
\end{equation}

The erased matrix fails to have full row rank precisely when
$z^{\mathsf T}\mathsf A_{\lambda,\mathsf E}=0$ for some nonzero
$z\in\mathbb F_2^e$. Extend $z^{\mathsf T}$ to a length-$N$ row
$w$ by putting its entries at the positions in $\mathsf E$ and
zeros elsewhere. Since $\mathsf A_\lambda$ consists of the first
$t_c$ columns of $\mathsf M_\lambda$,
\[
 z^{\mathsf T}\mathsf A_{\lambda,\mathsf E}=0
 \quad\Longleftrightarrow\quad
 (w\mathsf M_\lambda)_{1:t_c}=0 .
\]
There are $2^e-1$ such nonzero candidate rows $w$. For each one,
$w\mathsf M_\lambda$ is uniform over the nonzero rows, of which
$2^{N-t_c}-1$ vanish in their first $t_c$ coordinates. Hence
\begin{equation}
 \Pr_\lambda\{\rank\mathsf A_{\lambda,\mathsf E}<e\}
 \le\frac{(2^e-1)(2^{N-t_c}-1)}{2^N-1}
 <2^{e-t_c}.
 \label{sm:eq:helper-erased-rank-bound}
\end{equation}
When $e=0$, the row-rank test passes automatically.
Substituting $N=2m_c$ and $t_c=m_c-3$ in the two bounds yields
\[
 \Pr_\lambda\{\text{at least one rank test fails}\}
 <(2^2+2^3)2^{e-m_c}=12\,2^{e-m_c}.
\]

To average over the erasure patterns, interchange the two
finite sums:
\[
 \begin{aligned}
 \bar s
 =\frac{2^{-N}}{2^N-1}
   \sum_{\lambda\in\mathbb K^\times}
   \sum_{\mathsf E\subseteq[N]}
      \mathbf1_{\{\mathsf E\in\mathcal G_\lambda\}}=2^{-N}\sum_{\mathsf E\subseteq[N]}
      \Pr_\lambda\{\mathsf E\in\mathcal G_\lambda\}.
 \end{aligned}
\]
For $e\le m_c-4$, the preceding rank bound applies to every
fixed set of size $e$; for $e>m_c-4$, the decoder rejects.
Writing $b_e=2^{-N}\binom Ne$ gives
\begin{equation}
 \bar s\ge
 \sum_{e=0}^{m_c-4}b_e
 -12\sum_{e=0}^{m_c-4}b_e\,2^{e-m_c}
 >\frac12-5b_{m_c}.
 \label{sm:eq:helper-mean-success}
\end{equation}
For the last inequality, use symmetry of the binomial distribution
and $b_e\le b_{m_c}$:
\[
 \begin{aligned}
 \sum_{e=0}^{m_c-4}b_e
 &=\frac{1-b_{m_c}}2-b_{m_c-1}-b_{m_c-2}-b_{m_c-3}
 \ge\frac12-\frac72 b_{m_c},\\
 12\sum_{e=0}^{m_c-4}b_e\,2^{e-m_c}
 &\le12b_{m_c}2^{-m_c}(2^{m_c-3}-1)
 <\frac32 b_{m_c}.
 \end{aligned}
\]
The central probability satisfies
\[
 b_{m_c}
 =\frac2\pi\int_0^{\pi/2}\cos^{2m_c}x\,dx
 \le\frac2\pi\int_0^\infty e^{-m_cx^2}\,dx
 =\frac1{\sqrt{\pi m_c}}.
\]
For the equality, integration by parts gives
$I_m=(2m-1)I_{m-1}/(2m)$ and $I_0=\pi/2$, where
$I_m=\int_0^{\pi/2}\cos^{2m}x\,dx$. For the inequality,
$(\ln\cos x+x^2/2)'=x-\tan x\le0$ on $[0,\pi/2)$, so
$\cos^{2m_c}x\le e^{-m_cx^2}$.

\emph{Choice of parameters.}
Writing $t_c=m_c-a$ gives the coefficient $2^{5-a}+2^a$ in
the combined rank bound. Its minimum over integer $a$ is $12$,
attained at $a=2,3$; we take $a=3$. If the data acceptance
cutoff is $e\le m_c-c$, the same two binomial estimates give
the loss coefficient
\[
 c-\tfrac12+12\,2^{1-c}
\]
multiplying $b_{m_c}$. The choice $c=4$ gives $5$, compared with
$11/2$ and $21/4$ for $c=3$ and $5$, respectively. At the
accepted boundary $e=m_c-4$, the rank-failure probability is
less than $3/4$.

Choose
\begin{equation}
 \begin{gathered}
 m_c=32\,768,\qquad N=65\,536,\qquad t_c=32\,765,\\
 r_c=26,\qquad L=(r_c+1)N=1\,769\,472.
 \end{gathered}
 \label{sm:eq:helper-explicit-parameters}
\end{equation}
Since $\pi>25/8$, these choices give $\pi m_c>320^2$ and
$b_{m_c}<1/320$. Equation~\eqref{sm:eq:helper-mean-success}
then gives $\bar s>31/64$. The header repetition gives
$N2^{-r_c}=2^{-10}$, so $q_H\ge1023/1024$. Therefore
\begin{equation}
 s>\frac{1023}{1024}\frac{31}{64}
   =\frac{31713}{65536},\qquad
 p=1-s<\frac{33823}{65536}=p_0 .
 \label{sm:eq:helper-failure}
\end{equation}
Conversely, acceptance requires $e\le m_c-4$, so
\[
 s\le\sum_{e=0}^{m_c-4}b_e<\frac12
\]
by binomial symmetry. Hence $p>1/2$. Together with the
reference--output identity~\eqref{sm:eq:helper-reference-identity},
this proves Eq.~\eqref{sm:eq:theorem-two-logical}.
\end{proof}

\emph{An explicit field representation.}
For the value $N=65\,536$ above, the basis used to compute
$\mathsf M_\lambda$ can also be specified without a search.
Set $\mathbb K_0=\mathbb F_2$, $c_0=1$, and for $j=1,\ldots,16$
define
\[
 \mathbb K_j
 =\mathbb K_{j-1}[z_j]/(z_j^2+z_j+c_{j-1}),
 \qquad c_j=c_{j-1}z_j .
\]
The absolute trace of $c_0$ is one. If $c_{j-1}$ has absolute
trace one, the displayed quadratic has no root in $\mathbb K_{j-1}$,
since every $x^2+x$ has absolute trace zero. The quadratic is
therefore irreducible. Its two roots differ by one, giving
\[
 \operatorname{Tr}_{\mathbb K_j/\mathbb K_{j-1}}(z_j)=1,
 \qquad
 \operatorname{Tr}_{\mathbb K_j/\mathbb K_{j-1}}(c_j)=c_{j-1}.
\]
Transitivity of the trace shows that $c_j$ again has absolute
trace one. Thus $\mathbb K_{16}$ has degree $2^{16}=N$ over
$\mathbb F_2$ and may be taken as $\mathbb K$. Order the basis
monomials $z_1^{\epsilon_1}\cdots z_{16}^{\epsilon_{16}}$
lexicographically by
$(\epsilon_1,\ldots,\epsilon_{16})\in\{0,1\}^{16}$.
The reductions $z_j^2=z_j+c_{j-1}$ then determine every column
$[\lambda\beta_i]$ of $\mathsf M_\lambda$.

\emph{A shorter code with a fixed label.}
The data success probabilities $s_\lambda=2^{-N}|\mathcal G_\lambda|$
have average $\bar s>31/64$. Hence some label $\lambda_*$ satisfies
$s_{\lambda_*}>31/64$. If Alice and Bob fix such a label in
advance, they can omit the header and use $V_{\lambda_*}$ on
the $N$ data positions. Bob tests
$\mathsf E\in\mathcal G_{\lambda_*}$ and applies the same recovery
on acceptance; otherwise he returns the logical erasure flag.
Equation~\eqref{sm:eq:helper-recovery-factorization} gives the
exact logical channel $\mathcal E_{p_{\rm fix}}^{(32)}$, with
$p_{\rm fix}=1-s_{\lambda_*}$. Again $s_{\lambda_*}<1/2$ because
acceptance requires $e\le m_c-4$. Consequently,
\[
 \frac12<p_{\rm fix}<\frac{33}{64}<p_0 .
\]
Thus $N=65\,536$ uses suffice for a suitable fixed label.
The averaging argument does not identify that label; the
$L=1\,769\,472$ construction in the proposition requires no
selection of a good label.

Both erasure probabilities lie in the window of
Eq.~\eqref{sm:eq:helper-threshold}. Indeed,
$5^3<2^7$ gives $\hb(1/5)=\log_2 5-8/5<11/15$, and hence
\[
 p_0=\frac{33823}{65536}<\frac{12}{23}
 <\frac{4}{4+5\hb(1/5)}=p_* .
\]

\subsection{Positive joint capacity and environmental magic}

\begin{theorem}[Two convolution channels; Theorem 2 of the Letter]
\label{sm:thm:two-convolution}
Let $\N_{G,\sigma}$ be the $(7,7)$ channel of
Sec.~\ref{sm:sec:example}, or any of the seven-qudit channels
constructed in Theorem~\ref{sm:thm:crossratio} over $\F_d$.
Then
\begin{equation}
 \begin{gathered}
 Q(\N_{G,\sigma})=Q(\mathcal A_d)=P(\mathcal A_d)=0,\\
 Q(\N_{G,\sigma}\otimes\mathcal A_d)
 >\frac{8503}{55662776156160}>0 .
 \end{gathered}
 \label{sm:eq:theorem-two-rate}
\end{equation}
The explicit auxiliary code of
Proposition~\ref{sm:prop:auxiliary-simulation} gives a block input
with positive coherent information, from which the stated capacity
bound follows. The rate is in qubits per use of the pair: one use
of the seven-qudit channel $\N_{G,\sigma}$ and one use of the
one-qudit channel $\mathcal A_d$.

Both environment states, $\sigma$ and $\proj{\xi}$, are magic.
With the two interactions and the other environment held fixed,
replacing either environment by a stabilizer mixture gives
\begin{equation}
 \begin{aligned}
 Q(\N_{G,\eta}\otimes\mathcal A_d)&=0
       &&(\eta\in\STAB_7),\\
 Q(\N_{G,\sigma}\otimes\N_{G_{\rm aux},\zeta})&=0
       &&(\zeta\in\STAB_1).
 \end{aligned}
 \label{sm:eq:theorem-two-replacements}
\end{equation}
The two environments are prepared independently. The capacities
allow unrestricted joint encoding and decoding across both channels
and their repeated uses.
\end{theorem}

\begin{proof}
Let $\omega_{AC}$ be the equiprobable mixture of the two signal
purifications for $U_h$ in the proof of
Theorem~\ref{sm:thm:crossratio}. For the original coupling $G$,
the corresponding input is
\[
 \omega^G_{AC}
 =(S_{b/a}\otimes I_C)\,\omega_{AC}\,
  (S_{b/a}^\dagger\otimes I_C).
\]
For the explicit example in Sec.~\ref{sm:sec:example}, $a=b=1$,
so the input in Eq.~\eqref{sm:eq:smomega} is used without
rescaling. By Eqs.~\eqref{sm:eq:crossratio-normalization}
and~\eqref{sm:eq:crossratio-coherent-information}, sending $A$
through $\N_{G,\sigma}$ and $C$ through $\E_p^{(32)}$ gives
\[
 I_c(\omega^G,\N_{G,\sigma}\otimes\E_p^{(32)})
 =f(p)=\frac{1-p}{40}-\frac{p\,h_2(1/5)}{32}.
\]

To realize the erasure channel on $C$, apply the isometric encoder
of Proposition~\ref{sm:prop:auxiliary-simulation} to $C$ alone.
The physical input is
\[
 \widetilde\omega^G
 =(\operatorname{id}_A\otimes\mathcal V_d)(\omega^G).
\]
Alice sends $A$ through $\N_{G,\sigma}$ and the $L$ encoded qudits
through $\mathcal A_d^{\otimes L}$. Bob keeps the main output $B$
and applies $\mathcal D_d$ to the auxiliary outputs. By
Eq.~\eqref{sm:eq:theorem-two-logical}, the decoded channel on $C$
is exactly $\E_p^{(32)}$, with logical erasure probability
$1/2<p<p_0=33823/65536$. The coherent information after decoding
therefore satisfies
\begin{equation}
 I_c(\omega^G,\N_{G,\sigma}\otimes\E_p^{(32)})
 =f(p)
 >\frac{12-23p}{480}
 >\frac{8503}{31457280}>0 .
 \label{sm:eq:auxiliary-positive-information}
\end{equation}
Indeed, $5^3<2^7$ gives
$h_2(1/5)=\log_2 5-8/5<11/15$, and
$(12-23p_0)/480=8503/31457280$.

It remains to compare this value with the coherent information
before decoding. Let $R_0$ purify $\omega^G_{AC}$. Since the
encoder is an isometry, $R_0$ also purifies the encoded input
$\widetilde\omega^G$. Before decoding, Bob holds
$B_{\rm tot}=B B_h^{\otimes L}$; afterward he holds
$\widetilde B=B C_{\rm out}$, where
$C_{\rm out}=\widehat C\oplus\mathbb C\ket{\mathrm{er}}$ includes
both the recovered register and the erasure flag.
Represent Bob's operation by an isometry
$B_{\rm tot}\to\widetilde B D_0$ followed by a trace over $D_0$.
The coherent information is minus the conditional entropy of $R_0$
given Bob's output. Hence
\begin{equation}
 \begin{aligned}
 I_c(\widetilde\omega^G,
     \N_{G,\sigma}\otimes\mathcal A_d^{\otimes L})
 &=-S(R_0|B_{\rm tot})\\
 &=-S(R_0|\widetilde B D_0)\\
 &\ge -S(R_0|\widetilde B)=f(p).
 \end{aligned}
 \label{sm:eq:auxiliary-native-information}
\end{equation}
The inequality is strong subadditivity. The final equality holds
because the channel identity in Eq.~\eqref{sm:eq:theorem-two-logical}
also holds with the reference $R_0$ retained.

So far the input uses the private channel once and the auxiliary
$L$ times. For $L$ uses of the pair
$\N_{G,\sigma}\otimes\mathcal A_d$, place $\widetilde\omega^G$
on the first private input and all $L$ auxiliary inputs, and put
independent pure states on the remaining $L-1$ private inputs.
For each such pure input, the Stinespring output on $BK$ is pure,
so $S(B)=S(K)$ and its coherent information is zero. These
additional outputs are in tensor product with the output of
$\widetilde\omega^G$. Entropy additivity on product states and
the coding bound in Eq.~\eqref{sm:eq:model-coherent-information}
thus give
\[
 Q(\N_{G,\sigma}\otimes\mathcal A_d)
 \ge \frac{f(p)}{L}
 >\frac{8503}{31457280\cdot1769472}
 =\frac{8503}{55662776156160}.
\]
The private channel has zero quantum capacity by
Theorem~\ref{sm:thm:crossratio}. The auxiliary is self-complementary
on its entire input space by Eq.~\eqref{sm:eq:auxiliary-isometry},
so $P(\mathcal A_d)=Q(\mathcal A_d)=0$. This proves
Eq.~\eqref{sm:eq:theorem-two-rate}.

Theorem~\ref{sm:thm:crossratio} shows that $\sigma$ is magic.
A pure one-qudit stabilizer state has computational-basis support
of size $1$ or $d$~\cite{Gross2006}. The pure state $\ket{\xi}$
has support $\{1,-1\}$, so it too is magic.

First replace $\sigma$ by $\eta\in\STAB_7$.
By Eq.~\eqref{sm:eq:stabilizer-eb}, $\N_{G,\eta}$ is entanglement
breaking and hence antidegradable. The unchanged auxiliary
$\mathcal A_d$ is self-complementary. The tensor product of the
two antidegrading maps reconstructs the joint receiver output
from the joint complementary output. Thus
$Q(\N_{G,\eta}\otimes\mathcal A_d)=0$.

Now keep $\sigma$ fixed and replace $\proj{\xi}$ by
$\zeta\in\STAB_1$. By Eq.~\eqref{sm:eq:stabilizer-eb}, the new
auxiliary $\N_{G_{\rm aux},\zeta}$ is entanglement breaking, so
its Choi state is separable and in particular PPT. The unchanged
private channel also has a PPT Choi state. Up to reordering
tensor factors,
\[
 \bigl(J_{\N_{G,\sigma}\otimes\N_{G_{\rm aux},\zeta}}\bigr)^{T_{B B_h}}
 \simeq
 J_{\N_{G,\sigma}}^{T_B}
 \otimes J_{\N_{G_{\rm aux},\zeta}}^{T_{B_h}}
 \succeq 0 .
\]
The joint channel therefore has zero quantum capacity, proving
the second equality in Eq.~\eqref{sm:eq:theorem-two-replacements}.
\end{proof}

The auxiliary replacement can also be made across a whole block.
Let $\zeta_L\in\STAB_L$ be a joint state of the $L$ auxiliary
environment qudits, independent of the private environments.
With the tensor factors paired by position, the auxiliary block
channel is
\[
 \mathcal M_{\zeta_L}(X)
 =\Tr_{E_{h,1}\cdots E_{h,L}}\!\left[
   U_{\rm aux}^{\otimes L}(X\otimes\zeta_L)
   (U_{\rm aux}^{\otimes L})^\dagger\right].
\]
Equation~\eqref{sm:eq:stabilizer-eb} applies to this $L$-qudit
convolution, so $\mathcal M_{\zeta_L}$ is entanglement breaking
even when $\zeta_L$ is correlated across positions. Since
$\N_{G,\sigma}^{\otimes L}$ has a PPT Choi state, the same
tensor-product argument gives
\[
 Q\!\left(\N_{G,\sigma}^{\otimes L}\otimes
          \mathcal M_{\zeta_L}\right)=0 .
\]

\subsection{An environmental upper bound on the joint quantum capacity}

Let $\mathcal Z$ denote complete dephasing in the computational
basis,
\[
 \mathcal Z(X)
 =\sum_{x\in\F_d^n}\proj{x}\,X\,\proj{x}.
\]
We write $\mathcal Z_Y$ or $\mathcal Z_B$ when the register matters.
For states $\rho,\tau$, let
\[
 D_{\mathrm{rel}}(\rho\Vert\tau)
 =\Tr\!\left[\rho(\log_2\rho-\log_2\tau)\right]
\]
if $\operatorname{supp}\rho\subseteq\operatorname{supp}\tau$, and
set $D_{\mathrm{rel}}(\rho\Vert\tau)=+\infty$ otherwise.
The modified relative entropy of magic is
\[
 \operatorname{MRM}(\sigma)
 :=\min_{\tau\in\mathrm{MSPS}_n}D_{\mathrm{rel}}(\sigma\Vert\tau).
\]
Here $\mathrm{MSPS}_n$ is the set of minimal stabilizer-projection
states: normalized projectors onto joint eigenspaces of commuting
$n$-qudit Weyl operators~\cite{BGJ2026,XiongEtAl2026Transport}.
This set is contained in $\STAB_n$ but does not include all
stabilizer mixtures.

\begin{proposition}[Environmental upper bound]
\label{sm:prop:activated-mrm-bound}
Let $d$ be an odd prime, $G$ a positive BGJ coupling on $n$ qudits,
and $\sigma$ an arbitrary environment state. For every
finite-dimensional antidegradable auxiliary channel $\mathcal A$,
\begin{align}
 Q(\N_{G,\sigma}\otimes\mathcal A)
 &\le
 \min_{U\in\operatorname{Cliff}_n}
 \left\{S\!\left(\mathcal Z_Y(U\sigma U^\dagger)\right)-S(\sigma)\right\}
 \notag\\
 &\le \operatorname{MRM}(\sigma),
 \label{sm:eq:sm-activated-mrm-bound}
\end{align}
where $\operatorname{Cliff}_n$ is the group of Clifford unitaries
on the $n$ environment qudits.
\end{proposition}

\begin{proof}
\emph{Dephasing in the computational basis.}
Write
\[
 G^{-\mathsf T}=\begin{pmatrix}a&b\\c&e\end{pmatrix},
 \qquad
 U_G\ket{x,y}=\ket{ax+by,cx+ey}.
\]
Taking the partial trace over $E$ on a matrix unit gives
\begin{equation}
 \N_{G,\sigma}(\ket{x}\bra{x'})
 =\sum_{y,y'\in\F_d^n}
   \sigma_{yy'}\,\delta_{cx+ey,\,cx'+ey'}\,
   \ket{ax+by}\bra{ax'+by'} .
 \label{sm:eq:sm-activation-channel-expansion}
\end{equation}
If the environment is dephased first, then $y=y'$, and the
Kronecker delta forces $x=x'$ because $c\ne0$. If Bob's output
is dephased instead, the two surviving labels must satisfy
\[
 ax+by=ax'+by',\qquad cx+ey=cx'+ey'.
\]
Since $G^{-\mathsf T}$ is invertible, these equations again give
$x=x'$ and $y=y'$. Thus the two operations give the same channel:
\begin{equation}
 \N_{G,\mathcal Z_Y(\sigma)}
 =\mathcal Z_B\circ\N_{G,\sigma},
 \qquad
 \N_{G,\mathcal Z_Y(\sigma)}(\rho)
 =\sum_{x,y}\rho_{xx}\sigma_{yy}\proj{ax+by}.
 \label{sm:eq:sm-activation-dephasing-identity}
\end{equation}

This comparison channel is antidegradable. Indeed, a Kraus
representation is
\[
 L_{xy}=\sqrt{\sigma_{yy}}\,\ket{ax+by}\bra{x},
 \qquad \sum_{x,y}L_{xy}^\dagger L_{xy}=I_A .
\]
Measuring the complementary output in the Kraus-label basis gives
$(x,y)$ with probability $\sigma_{yy}\rho_{xx}$. Preparing
$\ket{ax+by}$ from this outcome reproduces Bob's state in
Eq.~\eqref{sm:eq:sm-activation-dephasing-identity}.

\emph{Comparison of the two joint channels.}
Set
\[
 \mathcal T=\N_{G,\sigma}\otimes\mathcal A,\qquad
 \overline{\mathcal T}
 =\N_{G,\mathcal Z_Y(\sigma)}\otimes\mathcal A .
\]
Fix a block length $k$ and any joint input $\rho_{A^k A'^k}$,
where $A'$ is the auxiliary input. Let $\psi_{RA^k A'^k}$ be a
purification, written as a density operator. Bob holds
$O=B^k B'^k$, where $B'$ is the auxiliary output. The two
reference--receiver states are
\[
 \Omega_{RO}
 =(\operatorname{id}_R\otimes\mathcal T^{\otimes k})(\psi),
 \qquad
 \overline\Omega_{RO}
 =(\operatorname{id}_R\otimes\overline{\mathcal T}^{\otimes k})(\psi).
\]
By Eq.~\eqref{sm:eq:sm-activation-dephasing-identity},
\[
 \overline\Omega_{RO}
 =(\operatorname{id}_R\otimes\mathcal Z_B^{\otimes k}
      \otimes\operatorname{id}_{B'^k})(\Omega_{RO}).
\]
Only the main receiver outputs are dephased.

Write $I(R\rangle O)_\Omega=S(\Omega_O)-S(\Omega_{RO})$.
Both factors of $\overline{\mathcal T}$ are antidegradable, so the
tensor product of their antidegrading maps reconstructs $O$ from
the full complementary output $E_{\rm tot}$, also for $k$ uses.
For the pure Stinespring output of the comparison channel, data
processing and purity therefore give
\[
 I(R\rangle O)_{\overline\Omega}
 \le I(R\rangle E_{\rm tot})
 =-I(R\rangle O)_{\overline\Omega}.
\]
In particular,
\begin{equation}
 I(R\rangle O)_{\overline\Omega}\le0.
 \label{sm:eq:sm-activation-comparison-nonpositive}
\end{equation}

For a pinching map $\mathcal P$,
\[
 D_{\mathrm{rel}}(\theta\Vert\mathcal P(\theta))
 =S(\mathcal P(\theta))-S(\theta).
\]
Indeed, $\operatorname{supp}\theta\subseteq
\operatorname{supp}\mathcal P(\theta)$, and
$\log_2\mathcal P(\theta)$ is block diagonal on its support.
Its expectation is therefore the same in $\theta$ and
$\mathcal P(\theta)$. Applying this identity to $RO$ and to $O$
gives
\begin{align}
 I(R\rangle O)_\Omega-I(R\rangle O)_{\overline\Omega}
 &=
 D_{\mathrm{rel}}(\Omega_{RO}\Vert\overline\Omega_{RO})
 -D_{\mathrm{rel}}(\Omega_O\Vert\overline\Omega_O)
 \notag\\
 &\le D_{\mathrm{rel}}(\Omega_{RO}\Vert\overline\Omega_{RO}).
 \label{sm:eq:sm-activation-pinching-bound}
\end{align}

The two experiments use the same signal input $\psi$, interaction
$U_G^{\otimes k}$, and auxiliary $\mathcal A^{\otimes k}$. With
these fixed, preparing $\psi$, applying the channels, and tracing
out $E^k$ defines a channel $\mathcal F_\psi:Y^k\to RO$ on the
main environment input. In particular,
\[
 \Omega_{RO}=\mathcal F_\psi(\sigma^{\otimes k}),\qquad
 \overline\Omega_{RO}
 =\mathcal F_\psi\bigl(\mathcal Z_Y(\sigma)^{\otimes k}\bigr).
\]
Relative-entropy data processing and additivity on product states
give
\begin{align}
 D_{\mathrm{rel}}(\Omega_{RO}\Vert\overline\Omega_{RO})
 &\le D_{\mathrm{rel}}\!\left(
     \sigma^{\otimes k}\middle\Vert
     \mathcal Z_Y(\sigma)^{\otimes k}\right)
 \notag\\
 &=k\,[S(\mathcal Z_Y(\sigma))-S(\sigma)].
 \label{sm:eq:sm-activation-environment-dp}
\end{align}
Combining these estimates for the chosen input yields
\[
 \begin{aligned}
 I_c(\rho,\mathcal T^{\otimes k})
 &= I(R\rangle O)_\Omega\\
 &\le I(R\rangle O)_\Omega-I(R\rangle O)_{\overline\Omega}\\
 &\le D_{\mathrm{rel}}(\Omega_{RO}\Vert\overline\Omega_{RO})\\
 &\le k\,[S(\mathcal Z_Y(\sigma))-S(\sigma)].
 \end{aligned}
\]
The first inequality uses
Eq.~\eqref{sm:eq:sm-activation-comparison-nonpositive}.
Since the input and block length were arbitrary, the regularized
coherent-information formula gives
\begin{equation}
 Q(\N_{G,\sigma}\otimes\mathcal A)
 \le S(\mathcal Z_Y(\sigma))-S(\sigma).
 \label{sm:eq:sm-activation-basis-bound}
\end{equation}

\emph{Dephasing in a stabilizer basis.}
To optimize Eq.~\eqref{sm:eq:sm-activation-basis-bound}, we need
the analogue of Eq.~\eqref{sm:eq:sm-activation-dephasing-identity}
in each stabilizer basis. It follows from the action of $U_G$ on
a single environmental Weyl operator; see also the Clifford
transport result of Ref.~\cite{XiongEtAl2026Transport}.

Let $z=(u,v)\in\F_d^n\oplus\F_d^n$, with $u$ the phase label
and $v$ the shift label of $W_Y(z)$. Put $\kappa=ae-bc$.
Writing the output labels as $\beta=ax+by$ and $\epsilon=cx+ey$,
we have
\[
 y=\kappa^{-1}(-c\beta+a\epsilon).
\]
An environmental phase $u\cdot y$ thus gives phase labels
$-c\kappa^{-1}u$ on $B$ and $a\kappa^{-1}u$ on $E$.
An environmental shift $y\mapsto y+v$ shifts the two outputs
by $bv$ and $ev$, respectively. Hence, up to an overall phase,
$U_G(I_A\otimes W_Y(z))U_G^\dagger$ is the product of two Weyl
operators, with Bob's label given by
\[
 \Lambda_B(u,v)=(-c\kappa^{-1}u,\,bv).
\]
The Weyl operator on $E$ drops out under the partial trace.
Consequently, for every input $\rho$,
\[
 \N_{G,W_Y(z)\sigma W_Y(z)^\dagger}(\rho)
 =W_B(\Lambda_Bz)\,\N_{G,\sigma}(\rho)\,
   W_B(\Lambda_Bz)^\dagger .
\]

Let $M$ be a Lagrangian subspace of the environmental phase space.
Dephasing in the joint eigenbasis of its Weyl operators is given by
\[
 \mathcal Z_M(X)=\frac1{d^n}\sum_{z\in M}W(z)XW(z)^\dagger .
\]
To verify this, consider a matrix unit between two joint
eigenvectors. Conjugation by $W(z)$ multiplies it by the ratio
of their eigenvalue characters. Averaging over $z\in M$ gives
one for the same character and zero for distinct characters.
Since $M$ is Lagrangian, each character labels a one-dimensional
eigenspace.

The map $\Lambda_B$ is invertible and satisfies
\[
 [\Lambda_B z,\Lambda_B z']
 =-\frac{bc}{\kappa}[z,z'].
\]
Thus $\Lambda_B M$ is also Lagrangian. Averaging the preceding
channel identity over $M$ gives
\[
 \N_{G,\mathcal Z_M(\sigma)}
 =\mathcal Z_{\Lambda_B M}\circ\N_{G,\sigma}.
\]
The channel on the right has outputs diagonal in a fixed
orthonormal basis. It is therefore a measure-and-prepare channel,
hence entanglement breaking and antidegradable. The comparison
in Eqs.~\eqref{sm:eq:sm-activation-comparison-nonpositive}--%
\eqref{sm:eq:sm-activation-environment-dp} now applies with
$\mathcal Z_M$ on $Y$ and $\mathcal Z_{\Lambda_B M}$ on $B$.
It gives
\[
 Q(\N_{G,\sigma}\otimes\mathcal A)
 \le S(\mathcal Z_M(\sigma))-S(\sigma)
 \qquad\text{for every Lagrangian }M .
\]
Every stabilizer basis is taken to the computational basis by
a Clifford unitary, and conversely. Thus minimizing over $M$
is equivalent to minimizing
$S(\mathcal Z_Y(U\sigma U^\dagger))-S(\sigma)$ over
$U\in\operatorname{Cliff}_n$. This proves the first inequality
in Eq.~\eqref{sm:eq:sm-activated-mrm-bound}.

For the second inequality, take any $\tau\in\mathrm{MSPS}_n$
with $D_{\mathrm{rel}}(\sigma\Vert\tau)<\infty$. The labels of
the commuting Weyl operators defining $\tau$ span an isotropic
subspace. Extend it to a Lagrangian subspace $M$. Then $\tau$
is diagonal in the associated stabilizer basis, so
$\mathcal Z_M(\tau)=\tau$. Since
$\operatorname{supp}\sigma\subseteq\operatorname{supp}\tau$,
the same support inclusion holds for $\mathcal Z_M(\sigma)$.
Taking the expectation of $\log_2\tau$ in this basis gives
\[
 \begin{aligned}
 D_{\mathrm{rel}}(\sigma\Vert\tau)
 &=D_{\mathrm{rel}}(\sigma\Vert\mathcal Z_M(\sigma))
   +D_{\mathrm{rel}}(\mathcal Z_M(\sigma)\Vert\tau)\\
 &\ge S(\mathcal Z_M(\sigma))-S(\sigma).
 \end{aligned}
\]
The minimum dephasing entropy increase over all stabilizer bases
is therefore no greater than $D_{\mathrm{rel}}(\sigma\Vert\tau)$
for any such $\tau$. Minimizing over $\tau\in\mathrm{MSPS}_n$
proves the second inequality.
\end{proof}

\emph{Evaluation for the $(7,7)$ environment.}
In Eq.~\eqref{sm:eq:simple-state}, the $256$ coherent pairs have
disjoint computational-basis supports. For each pair, the
corresponding block of $\sigma$ is
\[
 \frac1{640}
 \begin{pmatrix}1&H_{ij}\\ H_{ij}&1\end{pmatrix},
 \qquad H_{ij}\in\{1,-1\}.
\]
Its weight is $2/640=1/320$. After normalization, the block is
pure; computational-basis dephasing turns it into $I_2/2$ and
increases its entropy by one bit. The $32$ singleton blocks and
all block weights remain unchanged. By the entropy formula for
a direct sum,
\begin{equation}
 Q(\N_\sigma\otimes\mathcal A_7)
 \le S(\mathcal Z_Y(\sigma))-S(\sigma)
 =256\cdot\frac1{320}\cdot 1
 =\frac45 .
 \label{sm:eq:sm-activation-four-fifths}
\end{equation}
Here $\mathcal A_7$ is antidegradable on its full input space by
Eq.~\eqref{sm:eq:auxiliary-isometry}. The upper bound is in qubits
per use of the joint channel and is independent of the block
length used to prove achievability.

\subsection{Environmental variation and robustness}
\label{sm:sec:resource-control}

For the $(7,7)$ example, keep both interactions and the auxiliary
environment fixed. Vary only the main environment, setting
\begin{equation}
 \sigma_\mu=(1-\mu)\mathcal Z_Y(\sigma)+\mu\sigma,
 \qquad 0\le\mu\le1.
 \label{sm:eq:resource-state}
\end{equation}
All computational-basis populations remain fixed, while each
off-diagonal entry is multiplied by $\mu$.

\emph{The main channel remains PPT.}
Each coherent pair in $640\sigma_\mu$ has eigenvalues $1\pm\mu$.
The entry rearrangement in
Eqs.~\eqref{sm:eq:example-choi-blocks}
and~\eqref{sm:eq:simple-pt-block} gives
\begin{equation}
 640\mathcal R_3(\sigma_\mu)\simeq I_{512}\oplus
 \begin{pmatrix}
  4I_{16}&\mu H_{16}\\
  \mu H_{16}^{\mathsf T}&4I_{16}
 \end{pmatrix}\oplus0 .
 \label{sm:eq:resource-ppt}
\end{equation}
Since $H_{16}H_{16}^{\mathsf T}=16I_{16}$, the eigenvalues of
the middle block are $4\pm4\mu$. Thus the Choi state of
$\N_{\sigma_\mu}$ is PPT and $Q(\N_{\sigma_\mu})=0$ throughout
the path.

Along this path the joint quantum capacity is nondecreasing in
$\mu$. To see this, write
$\mathcal L_t=t\,\operatorname{id}_B+(1-t)\mathcal Z_B$.
Linearity in the environment and
Eq.~\eqref{sm:eq:sm-activation-dephasing-identity} give
\[
 \N_{\sigma_\mu}=\mathcal L_\mu\circ\N_\sigma .
\]
Since $\mathcal Z_B^2=\mathcal Z_B$, one has
$\mathcal L_s\circ\mathcal L_t=\mathcal L_{st}$. Consequently,
for $0\le\mu_1\le\mu_2$ with $\mu_2>0$,
\[
 \N_{\sigma_{\mu_1}}
 =\mathcal L_{\mu_1/\mu_2}\circ\N_{\sigma_{\mu_2}}.
\]
Bob can therefore obtain the output at $\mu_1$ from the output
at $\mu_2$ by further dephasing the main output alone. With the
same auxiliary $\mathcal A_7$, this implies
\[
 Q(\N_{\sigma_{\mu_1}}\otimes\mathcal A_7)
 \le Q(\N_{\sigma_{\mu_2}}\otimes\mathcal A_7).
\]

\emph{Private information and activation witnesses.}
For the two signal states in Eq.~\eqref{sm:eq:simple-state}, Bob's
conditional outputs are diagonal in the computational basis.
They are therefore unchanged by $\mathcal L_\mu$, and
$H(T|B)=h_2(1/5)/32$ as in the original example.

To compute Eve's conditional entropy, choose orthonormal vectors
$\ket{ij,0}_F,\ket{ij,1}_F$ for each pair and set
\[
 \ket{f^u_{ij}}=\ket{ij,0},\qquad
 \ket{f^v_{ij}}
 =H_{ij}\bigl(\mu\ket{ij,0}
              +\sqrt{1-\mu^2}\ket{ij,1}\bigr).
\]
In Eq.~\eqref{sm:eq:simple-purification}, replace the term for
pair $(i,j)$ by
\[
 \ket{u_{ij}}_Y\ket{f^u_{ij}}_F
 +\ket{v_{ij}}_Y\ket{f^v_{ij}}_F ,
\]
leaving the overall normalization and the singleton terms
unchanged. The flag spaces of distinct pairs and singletons are
mutually orthogonal. Within each pair,
$\langle f^v_{ij}|f^u_{ij}\rangle=\mu H_{ij}$, so tracing out
$F$ gives precisely the off-diagonal entry $\mu H_{ij}/640$ of
$\sigma_\mu$.

Consider the shared Eve sector for a fixed pair $(i,j)$.
The two signals give the same computational-basis label on $E$,
\[
 \epsilon_{ij}=u_{ij}-2r_i=v_{ij}-2s_j=-3r_i-3s_j.
\]
Conditional on this sector, the two letters are equiprobable,
and their states on $K=EF$ are
\[
 \ket{\epsilon_{ij}}_E\ket{f^u_{ij}}_F,\qquad
 \ket{\epsilon_{ij}}_E\ket{f^v_{ij}}_F .
\]
Their overlap has magnitude $\mu$. Their equal mixture thus has
nonzero spectrum $\{(1+\mu)/2,(1-\mu)/2\}$, with a zero
eigenvalue omitted at $\mu=1$. Since the two conditional states
are pure, the Holevo information in this sector is
$h_2((1+\mu)/2)$.

The sector probabilities depend only on the unchanged populations.
As in Sec.~\ref{sm:sec:example}, each of the $256$ shared sectors
has probability $1/10240$, and every other sector determines the
letter. Averaging the conditional entropy over these orthogonal
sectors gives
\begin{equation}
 H(T|B)=\frac{h_2(1/5)}{32},\qquad
 H(T|K)=\frac1{40}
 \left[1-h_2\!\left(\frac{1+\mu}{2}\right)\right].
 \label{sm:eq:resource-private}
\end{equation}
In particular, at $\mu=0$ the two flags in each shared sector
are orthogonal and Eve knows the letter. At $\mu=1$ they agree
up to a phase, recovering $H(T|K)=1/40$. No measurement of the
flags is assumed in Eq.~\eqref{sm:eq:resource-private}.

For the fixed input $\omega_{AC}$ of Eq.~\eqref{sm:eq:smomega},
the erasure identity~\eqref{sm:eq:sm-erasure-general} now yields
\begin{equation}
 \begin{aligned}
 F_p(\mu)
 &:= I_c(\omega,\N_{\sigma_\mu}\otimes\E_p^{(32)})\\
 &=\frac{1-p}{40}
   \left[1-h_2\!\left(\frac{1+\mu}{2}\right)\right]
   -\frac{p\,h_2(1/5)}{32}.
 \end{aligned}
 \label{sm:eq:resource-joint}
\end{equation}
At $\mu=1$ this is $f(p)$ from Eq.~\eqref{sm:eq:helper-fp}.
At $p=1/2$, twice $F_{1/2}(\mu)$ is the private information
$H(T|K)-H(T|B)$ of the chosen ensemble.

The explicit code of Proposition~\ref{sm:prop:auxiliary-simulation}
uses $L=1\,769\,472$ copies of $\mathcal A_7$ and has logical
erasure probability $p<p_0=33823/65536$. This probability is
independent of the main environment. Since $F_p(\mu)$ is strictly
decreasing in $p$, the coding argument of
Theorem~\ref{sm:thm:two-convolution} gives the lower bound
$\max\{0,F_{p_0}(\mu)\}/L$ per use of the physical channel pair.

For the upper bound, apply
Proposition~\ref{sm:prop:activated-mrm-bound} in the computational
basis. Each of the $256$ pair blocks has weight $2/640$.
Its normalized eigenvalues are $(1\pm\mu)/2$ before dephasing
and $(1/2,1/2)$ afterward. The singleton blocks and all block
weights remain unchanged. Thus
\[
 \frac{\max\{0,F_{p_0}(\mu)\}}{L}
 \;\le\; Q(\N_{\sigma_\mu}\otimes\mathcal A_7)
 \;\le\; \frac45
 \left[1-h_2\!\left(\frac{1+\mu}{2}\right)\right].
\]
The upper bound gives zero joint capacity at $\mu=0$ and recovers
Eq.~\eqref{sm:eq:sm-activation-four-fifths} at $\mu=1$.
A zero of the lower bound alone does not determine the capacity.

\emph{Environmental mana.}
Write $D=7^7$ and $\omega_7=e^{2\pi i/7}$. For
$q,\boldsymbol{k}\in\F_7^7$, use the discrete Wigner function
and mana~\cite{Gross2006,Veitch2014}
\[
 W_\eta(q,\boldsymbol{k})
 =\frac1D\sum_{z\in\F_7^7}
   \omega_7^{-\boldsymbol{k}\cdot z}
   \langle q+z/2|\eta|q-z/2\rangle,\qquad
 \mathcal M(\eta)
 =\log_2\sum_{q,\boldsymbol{k}}|W_\eta(q,\boldsymbol{k})|.
\]
The diagonal entries of $\sigma_\mu$ contribute nonnegative
Wigner values of total weight one. The off-diagonal entries of
pair $(i,j)$ contribute only at the midpoint
\[
 q_{ij}=\frac{u_{ij}+v_{ij}}2
       =-2r_i-2s_j=(-2,-2a_i,-2a_j).
\]
The last two coordinate blocks distinguish all $256$ midpoints.
Their first coordinate is $5$ in $\F_7$, whereas the first
coordinates of the diagonal support are $6,4,3,0$. Thus the
contribution at each midpoint comes from just one coherent pair.
With $z_{ij}=u_{ij}-v_{ij}\ne0$, it is
\[
 W_{\sigma_\mu}(q_{ij},\boldsymbol{k})
 =\frac{2\mu H_{ij}}{640D}
   \cos\!\left(\frac{2\pi}{7}\boldsymbol{k}\cdot z_{ij}\right).
\]
As $\boldsymbol{k}$ varies, the nonzero linear functional
$\boldsymbol{k}\mapsto\boldsymbol{k}\cdot z_{ij}$ takes each
value in $\F_7$ exactly $D/7$ times. The cosine takes both signs,
so $\sigma_\mu$ has a negative Wigner value whenever $\mu>0$.
Since every stabilizer mixture has a nonnegative Wigner function,
$\sigma_\mu$ is magic for every $\mu>0$. At $\mu=0$ it is a
mixture of computational-basis states. Summing the absolute
values gives
\begin{equation}
 \mathcal M(\sigma_\mu)=\log_2(1+c_7\mu),\qquad
 c_7=\frac4{35}\sum_{j=0}^{6}
     \left|\cos\!\left(\frac{2\pi j}{7}\right)\right|.
 \label{sm:eq:resource-mana}
\end{equation}

Figure~\ref{sm:fig:resource-control} plots the two witnesses
against this mana. The fixed-ensemble private information vanishes
at the unique $\mu_* \in (0,1)$ satisfying
\[
 h_2\!\left(\frac{1+\mu_*}{2}\right)
 =1-\frac54 h_2(1/5).
\]
It is positive for $\mu>\mu_*=0.974795\ldots$.
The conservative witness $F_{p_0}$ crosses zero at
$\mu=0.992018\ldots$. At this second zero the actual code already
has $F_p(\mu)>0$, since $p<p_0$. Neither zero determines the
onset of positive quantum capacity for the channel pair.

\begin{figure}[t]
 \centering
 \includegraphics[width=.9\linewidth]{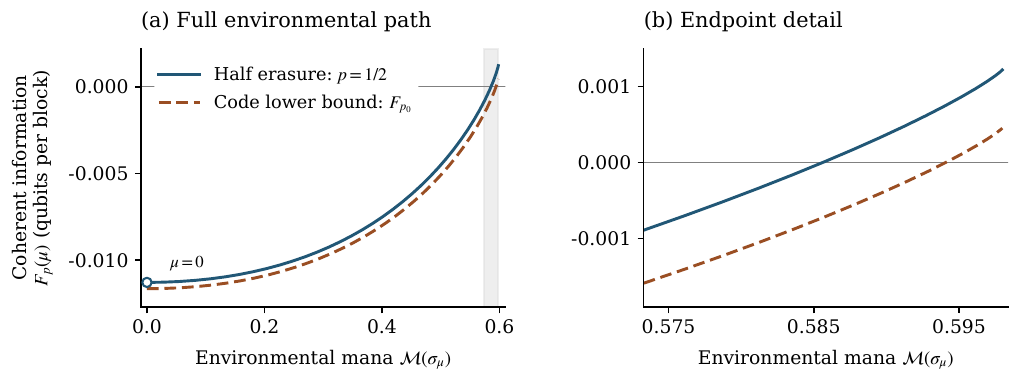}
 \caption{Activation witnesses for the $(7,7)$ example as a
 function of the main environment's mana. The solid curve is
 $F_{1/2}(\mu)$ for the ideal half-erasure auxiliary; twice its
 value is the private information of the chosen ensemble. The
 dashed curve is $F_{p_0}(\mu)$, evaluated at the upper bound
 $p_0=33823/65536$ on the explicit code's erasure probability.
 Its positive values divided by $L=1\,769\,472$ give a rate
 lower bound in qubits per use of the physical channel pair.
 Panel (b) enlarges the shaded interval $0.95\le\mu\le1$ of
 panel (a). The authors used ChatGPT to help write the plotting code.}
 \label{sm:fig:resource-control}
\end{figure}

\emph{Robustness under perturbations of the environment.}
Keep the main interaction fixed and now allow perturbations in
arbitrary Hermitian directions. For $0<\delta<1$, set
\[
 \eta_\delta=(1-\delta)\sigma+\delta I_D/D.
\]
Since $\mathcal R_3(I_D)=I_D$ and
$\mathcal R_3(\sigma)\succeq0$, both matrices satisfy
\[
 \eta_\delta\succeq\frac{\delta}{D}I_D,\qquad
 \mathcal R_3(\eta_\delta)\succeq\frac{\delta}{D}I_D.
\]

We use the same signal ensemble, activation input, and auxiliary
code as at $\eta=\sigma$. For any fixed input $\rho$ with
purification $\varphi_{RA}$, written as a density operator,
\[
 I_c(\rho,\N_\eta)
 = S(\N_\eta(\rho))
   -S\!\left((\operatorname{id}_R\otimes\N_\eta)(\varphi_{RA})\right).
\]
Both states on the right depend linearly on $\eta$, so their
entropies are continuous in $\eta$. The same argument applies
to a fixed joint input when a fixed auxiliary channel is present.
For the private witness, Eq.~\eqref{sm:eq:model-holevo} gives
\[
 \Delta(\eta)
 = I_c(\bar\rho,\N_\eta)
   -\tfrac12\sum_{t=0}^1 I_c(\rho_t,\N_\eta),
 \qquad \bar\rho=\tfrac12(\rho_0+\rho_1).
\]
Thus $\Delta(\eta)$ and the two coherent-information witnesses
\[
 I_c(\omega,\N_\eta\otimes\E_{1/2}^{(32)}),\qquad
 I_c(\omega,\N_\eta\otimes\E_p^{(32)})
\]
are continuous. In the second expression, $p$ is the actual
logical erasure probability of the fixed convolution code.
All three quantities are strictly positive at $\eta=\sigma$,
and hence at $\eta_\delta$ for all sufficiently small $\delta>0$.

Fix such a $\delta$. For any traceless Hermitian perturbation
$X$, the entry rearrangement $\mathcal R_3$ preserves Hermiticity
and the Hilbert--Schmidt norm. Hence
\[
 \|X\|_\infty\le\|X\|_{\mathrm{HS}},\qquad
 \|\mathcal R_3(X)\|_\infty
 \le\|\mathcal R_3(X)\|_{\mathrm{HS}}
 =\|X\|_{\mathrm{HS}} .
\]
If $\|X\|_{\mathrm{HS}}<\delta/D$, then both $\eta_\delta+X$ and
$\mathcal R_3(\eta_\delta+X)$ are positive definite. The former
is a density operator, and the latter ensures that its channel
has a PPT Choi state. Shrinking this ball if necessary keeps
all three witnesses positive. Thus a neighborhood of
$\eta_\delta$ in the trace-one Hermitian matrices consists of
environments for which $P(\N_\eta)>0$, $Q(\N_\eta)=0$, and
both the half-erasure and convolution auxiliaries give positive
joint quantum capacity. Such neighborhoods occur arbitrarily
close to the original environment $\sigma$.

\paragraph*{Verification and computational tools.}
The proofs above are analytical. The accompanying supplemental code
tests small instances of the encoder and decoder on both diagonal
and off-diagonal logical matrix units. Separate
scripts reconstruct the data in Fig.~\ref{sm:fig:resource-control} from
the channel outputs and compare the curves in the PDF with their analytical
expressions. These checks do not simulate the full auxiliary code.
OpenAI ChatGPT assisted with the
constructions, algebraic checks, and preparation of these scripts;
the full AI Use Disclosure appears in the Letter.

\ifPrintLetter
\else
\bibliographystyle{apsrev4-2}
\bibliography{references}
\fi
\fi
\end{document}